\documentclass[twocolumn]{aastex701}

\usepackage{multirow}
\usepackage{booktabs}
\usepackage{natbib}
\usepackage{graphicx}
\usepackage{amsmath}
\usepackage{textgreek}
\usepackage{graphicx}
\usepackage{soul}
\usepackage{subcaption}
\usepackage{array}
\usepackage{tabularx}
\usepackage{float}
\usepackage{makecell}
\usepackage{color}
\usepackage{hyperref}

\begin{document}
\title{Dwarf and Intermediate-Mass Galaxies in MaNGA: Evidence for Different Evolutionary Trends}

\author[orcid=0009-0005-2923-9933,gname='Chandan',sname='Watts']{Chandan Watts}
\affiliation{Indian Institute of Astrophysics,II Block, Koramangala, Bengaluru 560 034, INDIA.}
\affiliation{Pondicherry University, R.V. Nagar, Kalapet, 605014, Puducherry, India}
\email[show]{chandan@iiap.res.in, chandanwatts510@gmail.com}  

\author[orcid=0009-0004-8898-5138]{Gothai L}
\affiliation{Indian Institute of Astrophysics,II Block, Koramangala, Bengaluru 560 034, INDIA.}
\affiliation{Pondicherry University, R.V. Nagar, Kalapet, 605014, Puducherry, India}
\email{gothai@gmail.com}

\author[orcid=0000-0002-3927-5402]{Sudhanshu Barway}
\affiliation{Indian Institute of Astrophysics,II Block, Koramangala, Bengaluru 560 034, INDIA.}
\affiliation{Pondicherry University, R.V. Nagar, Kalapet, 605014, Puducherry, India}
\email{sudhanshu.barway@iiap.res.in}

\begin{abstract}
We investigate the interplay between morphology, specific star formation rate (sSFR), and local environment using a sample of 7,408 galaxies from the SDSS-IV MaNGA survey. Our analysis spans stellar masses from dwarf to massive galaxies, enabling a unified view of how stellar mass and environment regulate galaxy evolution. Galaxies are classified by morphology (ellipticals (E), lenticulars (S0s), early-type spirals (ETS), and late-type spirals (LTS)) and local environmental density, with star formation activity traced using sSFR. Low-mass galaxies ($\log (M_{\star}/M_{\odot}) < 10$) are predominantly star-forming and dominated by LTS, whereas high-mass galaxies ($\log (M_{\star}/M_{\odot}) \geq 10$) are dominated by ETS and are largely quenched. By separating dwarf ($\log (M_{\star}/M_{\odot}) \leq 9.5$) and intermediate-mass galaxies ($9.5 < \log (M_{\star}/M_{\odot}) < 10$), we find that dwarf galaxies remain predominantly star-forming with only weak environmental dependence, whereas intermediate-mass galaxies exhibit clearer environmental trends toward quenching. Using the D4000 index as a tracer of long-term stellar population aging, we further show that dwarf E and S0s host systematically younger stellar populations than their intermediate-mass counterparts, implying reduced quenching efficiency and more gradual environmental processing in the dwarf regime. This distinction is not evident among spiral galaxies, whose stellar population properties are comparatively insensitive to the dwarf versus non-dwarf classification. Overall, these results indicate that the commonly defined low-mass galaxy population is not homogeneous and that dwarf and intermediate-mass galaxies show systematically different evolutionary trends. Treating them separately is therefore essential for interpreting galaxy evolution in the low-mass regime.
\end{abstract}


\section{Introduction} \label{sec:intro}
Galaxy evolution is governed by a complex interplay between internal and external processes that regulate morphology, star formation, and stellar populations. Internal mechanisms such as gas inflows, central starbursts, and feedback from active galactic nuclei (AGN) operate alongside environmental factors including galaxy interactions, mergers, and local density to shape a galaxy’s evolutionary pathway \citep[e.g.,][]{2014ARA&A..52..291C, 2017MNRAS.471.2687B, Gu_2021}. Understanding how these processes collectively drive star formation and morphological transformation across stellar mass regimes remains a central goal in extragalactic astronomy.

Galaxy morphology provides a key diagnostic of evolutionary state. Elliptical and lenticular galaxies typically host quenched, old stellar populations, while spiral galaxies span a broad range of star formation rates (SFRs) depending on disk structure and gas content \citep{Blanton_2009, Kormendy_2016}. Stellar mass and environment further modulate these trends: massive galaxies ($\log (M_{\star}/M_{\odot}) \gtrsim 10$) preferentially reside in denser regions and contain older stellar populations, whereas lower-mass systems are more actively star-forming and are more commonly found in lower-density environments \citep{2010ApJ...721..193P, 2018ApJ...853..155D, Kawinwanichakij_2017}.

Stellar mass acts as a fundamental parameter that links these processes. The correlation between stellar mass and SFR, known as the star-forming main sequence (SFMS), demonstrates that star formation is strongly mass-regulated \citep{Noeske_2007, Daddi_2007, Elbaz_2007}. Departures from the SFMS, such as starbursts or quenched systems, mark key evolutionary transitions driven by feedback, gas depletion, or environmental effects.

At the low-mass end, however, the interpretation of mass-regulated star formation is complicated by the inconsistent use of terminology in the literature. The terms \emph{low-mass} and \emph{dwarf} galaxies are often used interchangeably, despite the absence of a uniform stellar-mass boundary. While dwarf galaxies are commonly defined using thresholds in the range $\log (M_{\star}/M_{\odot}) \lesssim 9$--9.5, galaxies with stellar masses extending up to $\log (M_{\star}/M_{\odot}) \lesssim 10$ are frequently grouped together as a broader low-mass population \citep{2023ApJ...955L..18L, 2023MNRAS.518..724S, 2024MNRAS.528.5252M, 2024MNRAS.532..613B, 2025MNRAS.536..295M}. This practice can obscure physically meaningful distinctions in galaxy evolution. In this work, we explicitly separate dwarf galaxies from more massive low-mass galaxies and demonstrate that several trends commonly attributed to low-mass galaxies are largely dominated by the dwarf population.

In this context, we analyze 7,408 galaxies from the Mapping Nearby Galaxies at Apache Point Observatory (MaNGA) survey, spanning stellar masses from dwarfs to massive systems. Using morphology, specific star formation rate (sSFR), and local environmental density, we examine how these properties are coupled across different mass regimes. Our primary aim is to assess whether evolutionary trends observed in the low-mass regime represent generic behavior or are dominated by dwarf galaxies.

This paper is organized as follows. Section~\ref{sec:sample} describes the data and sample construction. Section~\ref{sec:results} presents our results across stellar mass and environmental regimes. Section~\ref{sec:conclusion} summarizes our main findings and discusses their implications for galaxy evolution and upcoming integral-field spectroscopic surveys.

\begin{figure}[htbp]
    \centering
    \includegraphics[width=1\columnwidth]{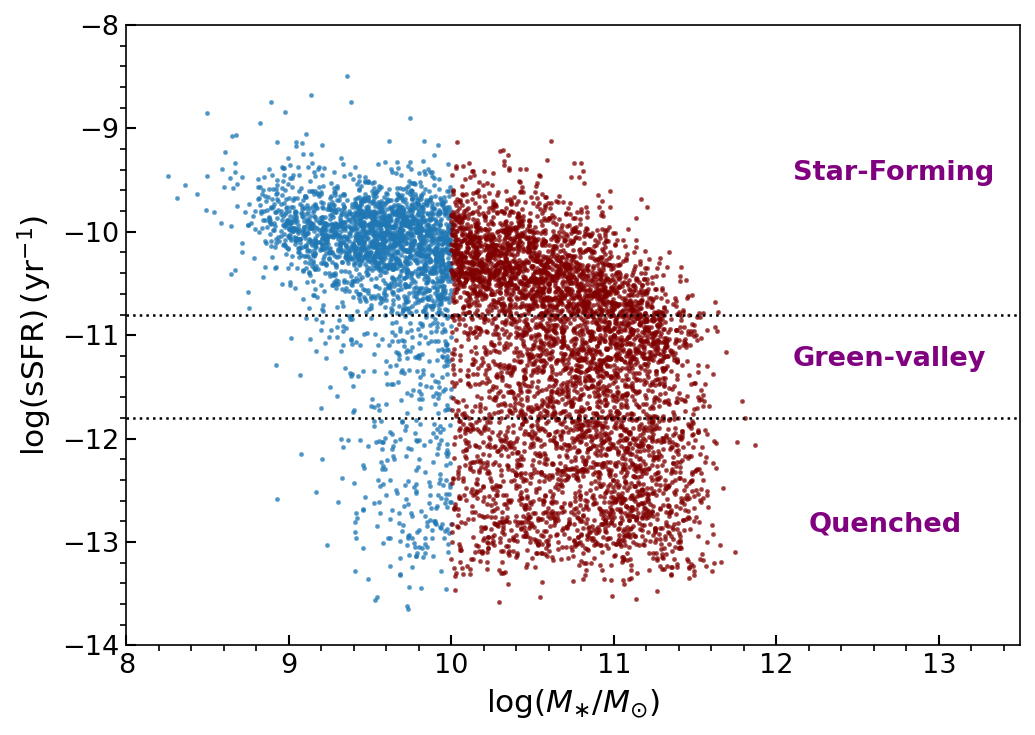}
    \caption{Scatter plot showing the distribution of galaxies as a function of stellar mass and specific star formation rate. High-mass galaxies ($\log (M_{\star}/M_{\odot}) \geq 10$) are shown in purple, while low-mass galaxies ($\log (M_{\star}/M_{\odot}) < 10$) are shown in blue.}
    \label{fig:sSFR_scatter_full_sample}
\end{figure}

\section{Data and Sample Description} \label{sec:sample}
Our study combines data from four major catalogs to construct the galaxy sample: the MaNGA survey from SDSS Data Release 17 (DR17; \cite{2015ApJ...798....7B, 2017AJ....154...86W}), the MaNGA Visual Morphology Catalog \citep{2022MNRAS.512.2222V}, the GALEX--SDSS--WISE Legacy Catalog (GSWLC; \cite{2016ApJS..227....2S, 2018ApJ...859...11S}), and the Environmental Density Catalog \citep{2006MNRAS.373..469B}. Below, we briefly describe each dataset and summarize the criteria used to construct the final sample.

The MaNGA survey, part of the fourth-generation Sloan Digital Sky Survey (SDSS-IV), uses integral field units (IFUs) composed of tightly packed optical fibre bundles to investigate the stellar and gaseous properties of nearby galaxies. The survey includes 11,273 galaxies spanning a wavelength range of 360--1000~nm, with a median redshift of $z \sim 0.03$ and typical stellar masses above $10^{9}~M_{\odot}$.

Morphological classifications are obtained by cross-matching the MaNGA DR17 catalog with the MaNGA Visual Morphology Catalog, which contains visual classifications for 10,126 galaxies, resulting in 10,125 unique MaNGA galaxies with assigned morphologies. These classifications are based on SDSS and DESI Legacy Survey imaging. Spiral galaxies are subdivided into Sa, Sab, Sb, Sbc, Sc, Scd, Sdm, Sm, Im, and S types, where the Sdm--S classes include very loosely wound spirals, irregular spirals, blue compact dwarfs (BCDs), barred irregulars (IrrB), irregular galaxies (Irr), dwarf spheroidals (dSph), and spiral mergers. Lenticular galaxies are classified into S0 and S0/a subtypes. 

For the purposes of this study, galaxies are grouped into four broad morphological classes: Ellipticals (E), Lenticulars (S0s; S0 and S0/a), Early-Type Spirals (ETS; Sa, Sab, Sb, Sbc), and Late-Type Spirals (LTS; Sc, Scd, Sd, Sdm, Sm, Im, and S).

SFRs and stellar masses are obtained from the GSWLC-X2 catalog. GSWLC covers galaxies within the GALEX footprint, encompassing $\sim$90\% of the SDSS sample, and provides SFR estimates derived from spectral energy distribution (SED) fitting across ultraviolet (FUV and NUV), optical ($u$, $g$, $r$, $i$, $z$), and mid-infrared (WISE: W1--W4) bands. The catalog contains 659,229 galaxies spanning $0.01 < z < 0.30$ with $r$-band magnitudes brighter than 18.

Cross-matching the MaNGA DR17 sample with GSWLC-X2 yields 8,536 galaxies with reliable stellar mass and SFR measurements. Local environmental density information is obtained from the Environmental Density Catalog \citep{2006MNRAS.373..469B}, ensuring that all galaxies in the sample have consistent measurements of morphology, star formation, and environment. After applying all selection criteria, the final master sample comprises 7,408 galaxies.

Galaxies are divided into three environmental regimes following \citet{2006MNRAS.373..469B}: low-density environments ($\log \Sigma \, (\mathrm{Mpc^{-2}}) \leq -0.5$), intermediate-density environments ($-0.5 < \log \Sigma \, (\mathrm{Mpc^{-2}}) < 0.5$), and high-density environments ($\log \Sigma \, (\mathrm{Mpc^{-2}}) \geq 0.5$). The sSFR, defined as the star formation rate normalized by stellar mass, serves as a tracer of recent star formation. Following the classification scheme of \citet{2014SerAJ.189....1S}, galaxies are divided on the basis of sSFR into star-forming ($\log(\mathrm{sSFR}) \geq -10.8$), green valley ($-10.8 < \log(\mathrm{sSFR}) < -11.8$), and quenched ($\log(\mathrm{sSFR}) \leq -11.8$) populations.

We use measurements of the D4000 index at the effective radius ($R_e$) from the MaNGA Pipe3D catalog \citep{2022NewA...9701895L,2022ApJS..262...36S}. The D4000 index provides a robust tracer of the luminosity-weighted stellar population age on gigayear timescales. Unlike the sSFR, which is sensitive to recent star formation, D4000 allows us to distinguish galaxies that have experienced long-term quenching from those that have only recently reduced their star formation activity.

The master sample is initially divided into two stellar mass regimes: 2,515 low-mass galaxies with $\log (M_{\star}/M_{\odot}) < 10$ and 4,893 high-mass galaxies with $\log (M_{\star}/M_{\odot}) \geq 10$, as illustrated in Figure~\ref{fig:sSFR_scatter_full_sample}. Throughout this paper, we further subdivide the low-mass regime by explicitly separating dwarf galaxies ($\log (M_{\star}/M_{\odot}) \leq 9.5$) from intermediate-mass galaxies ($9.5 < \log (M_{\star}/M_{\odot}) < 10$). Dwarf galaxies are defined using a stellar mass threshold of $\log (M_{\star}/M_{\odot}) \leq 9.5$ \citep{2023ApJ...955L..18L}, yielding a final sample of 977 galaxies. We verify their absolute $r$-band magnitudes ($M_r$) using the NASA--Sloan Atlas (NSA) catalog \citep{2020ApJ...900...56S} and find that all galaxies have $M_r > -20$. This is consistent with the dwarf galaxy definition adopted by \citet{2024MNRAS.528.5252M}, who similarly report $M_r > -20$ without imposing an explicit magnitude cut.

This distinction allows us to assess whether trends commonly attributed to low-mass galaxies are representative of the entire low-mass population or are driven primarily by dwarfs or intermediate-mass galaxies. This mass-based framework enables a systematic investigation of how morphology, sSFR, and local environmental density vary across stellar mass regimes. We use the resulting sample to explore the coupled roles of stellar mass and environment in regulating galaxy evolution.

\section{Results and Discussion}\label{sec:results}

We adopt a statistical approach to examine how galaxy properties vary across stellar mass regimes, with the primary goal of assessing whether trends commonly attributed to low-mass galaxies represent generic behavior or are dominated by specific subpopulations. 

\begin{figure*}[htbp]
    \centering
    \begin{subfigure}[b]{0.499\textwidth}
         \centering
         \includegraphics[width=\textwidth]{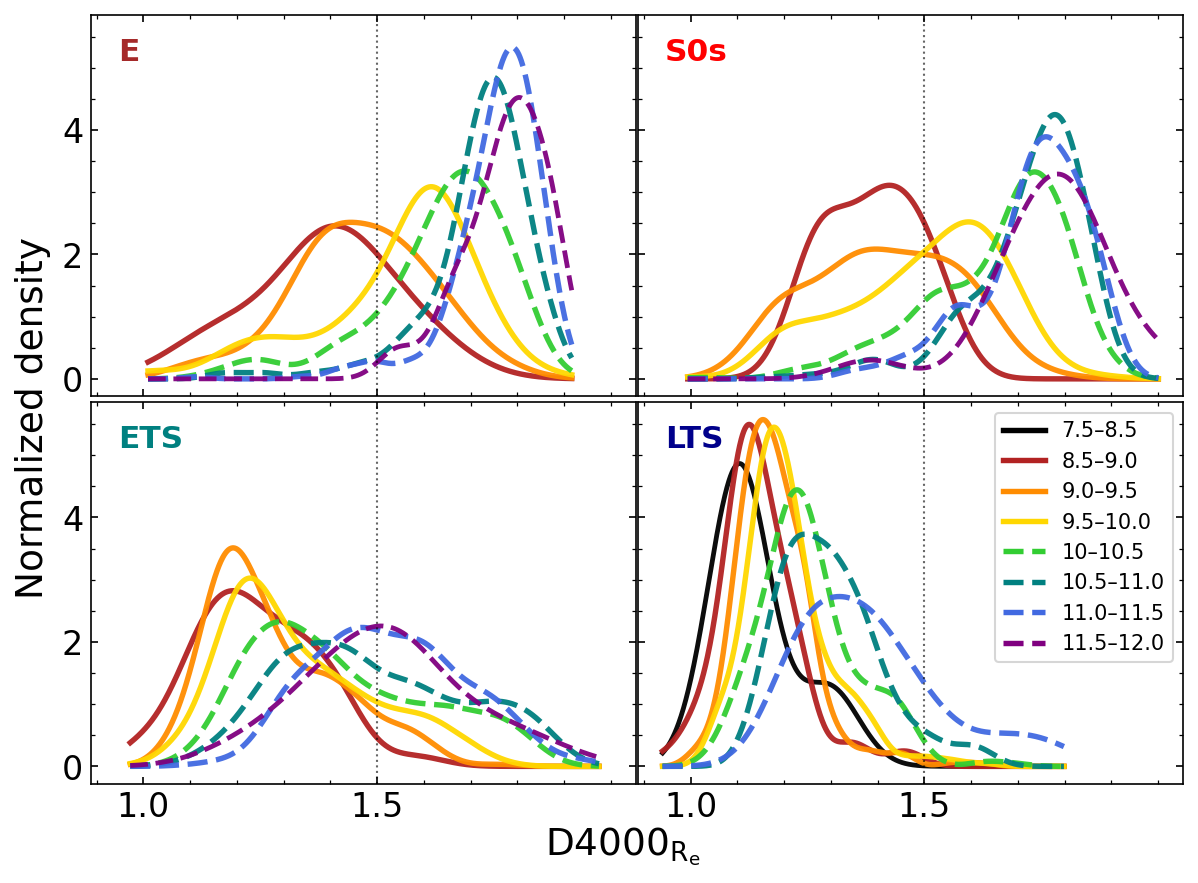}
     \end{subfigure}
     %\hfill
     \begin{subfigure}[b]{0.495\textwidth}
         \centering
         \includegraphics[width=\textwidth]{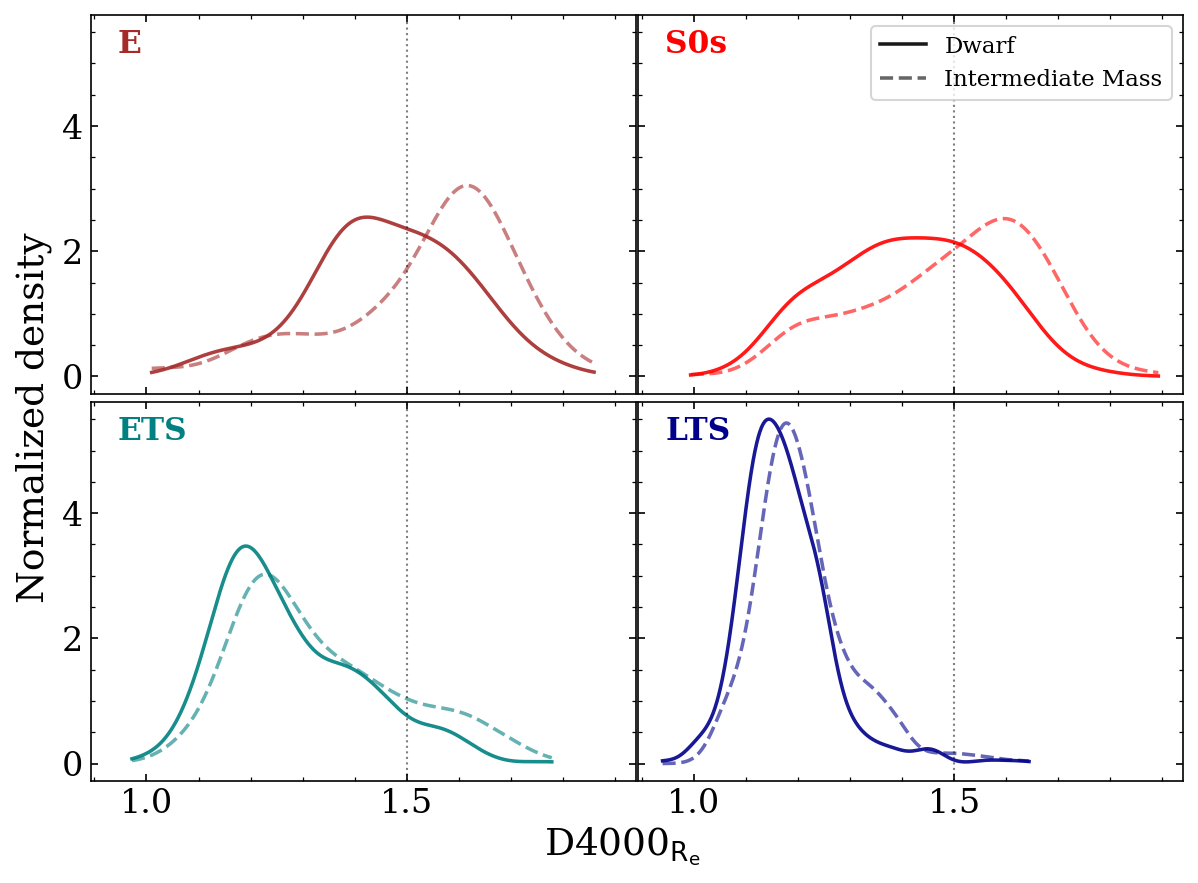}
     \end{subfigure}
    \caption{\textit{(Left)} Kernel density distribution of the D4000$_{R_e}$ index for galaxies across different stellar mass ranges and morphological types, spanning $\log(M_\ast/M_\odot)=7.5$--12.0, using a bin width of 0.5 $M_\odot$. \textit{(Right)} Kernel density distribution of the D4000$_{R_e}$ index for dwarf galaxies ($\log (M_{\star}/M_{\odot}) \leq 9.5$; solid lines), and intermediate-mass galaxies ($9.5 < \log (M_{\star}/M_{\odot}) < 10$; dotted lines) The horizontal dotted line at D4000$_{R_e}=1.5$ marks the approximate division between younger and older stellar populations.}
    \label{fig:all_bins}
\end{figure*}

We are primarily interested in low-mass galaxies, which remain relatively less explored in the literature. To investigate possible variations within this regime, we divide the full sample into stellar mass bins of 0.5 dex spanning $7.5 \leq \log(M_{\star}/M_{\odot}) \leq 12.0$, as shown in the left panel of Figure~\ref{fig:all_bins}. The figure presents the distribution of the D4000 index across these bins. The horizontal dotted line at D4000$_{R_e}=1.5$ marks the approximate division between younger and older stellar populations \citep{2003MNRAS.346.1055K, 2018ApJ...867..118K}.

This sample includes a substantial population of dwarf galaxies. In many observational studies, galaxies with log(M${\star}$/M${\odot}$) $<$ 10 are often grouped into a single low-mass population. However, because dwarf galaxies constitute a significant fraction of this population, their presence may strongly influence the overall trends.

Galaxies in the stellar mass bins with $\log(M_\star/M_\odot) \leq 9.5$ show broadly similar qualitative trends compared to those in the $9.5$–$10.0$ mass range. However, a clear quantitative differences are present between the dwarf galaxies ($\leq 9.5$) and the intermediate-mass galaxies ($9.5$–$10.0$) is evident for E and S0s, as shown in the right panel of the Figure~\ref{fig:all_bins}. To assess the statistical significance of this difference, we perform a Kolmogorov–Smirnov (KS) test. These results indicate that, while the overall trends are preserved across the low-mass regime, the strength of these trends varies systematically with stellar mass. The separation is strongest for E galaxies ($D=0.374$, $p=6.151\times10^{-5}$) and is also significant for S0s ($D=0.276$, $p=5.542\times10^{-6}$), indicating a pronounced shift toward older stellar populations at higher stellar masses. We explore these trends in greater detail in the following subsections.

To facilitate comparison with literature, we first summarize the distributions of morphology, star formation activity, and local environmental density for the full sample of galaxies with $\log (M_{\star}/M_{\odot}) < 10$.

\subsection{Low-mass galaxies} \label{sec:low_mass}
\begin{figure*}[htbp]
    \centering
    \begin{subfigure}[t]{0.47\textwidth}
        \includegraphics[width=\textwidth]{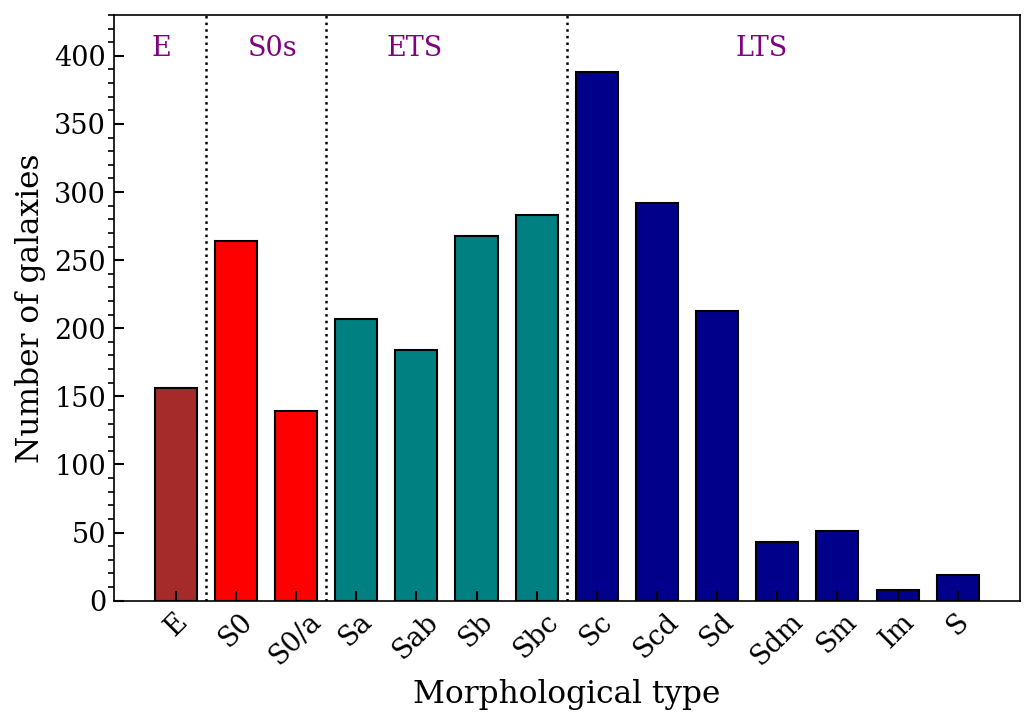} % Replace with your image file
        \caption{\scriptsize }
        \label{fig:morphological_all_histogram_low_mass}
    \end{subfigure}
    \begin{subfigure}[t]{0.47\textwidth}
        \includegraphics[width=\textwidth]{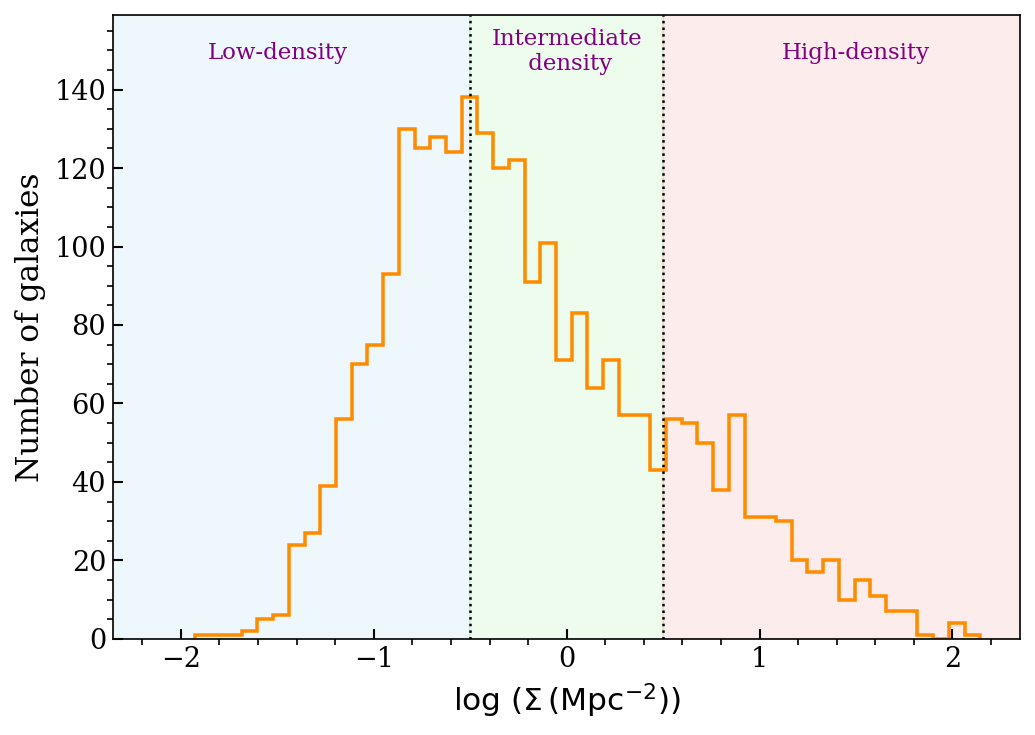}
        \caption{\scriptsize}
        \label{fig:density_histogram_low_mass}
    \end{subfigure}

    \vspace{0cm} % Space between rows

    \begin{subfigure}[t]{0.47\textwidth}
        \includegraphics[width=\textwidth]{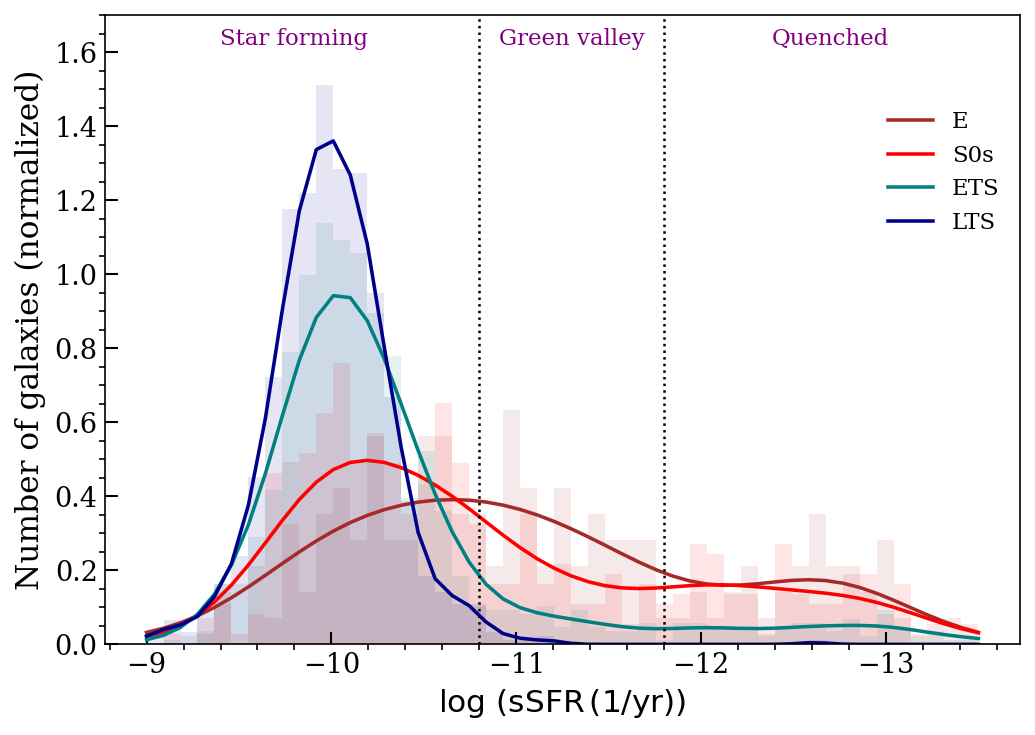}
        \caption{\scriptsize}
        \label{fig:sSFR_all_histogram_low_mass}
    \end{subfigure}
    \begin{subfigure}[t]{0.47\textwidth}
        \includegraphics[width=\textwidth]{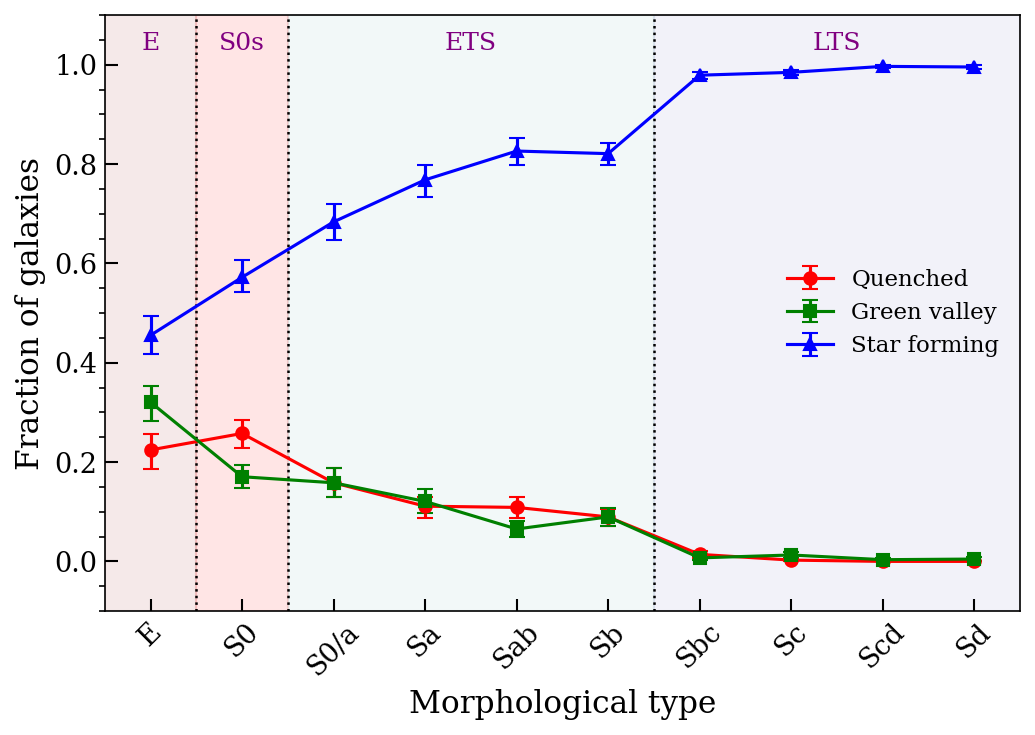}
        \caption{\scriptsize}
        \label{fig:morphology_all_fraction_low_mass}
    \end{subfigure}
    
    \vspace{0cm} % Space between rows
    
    \begin{subfigure}[t]{0.47\textwidth}
        \includegraphics[width=\textwidth]{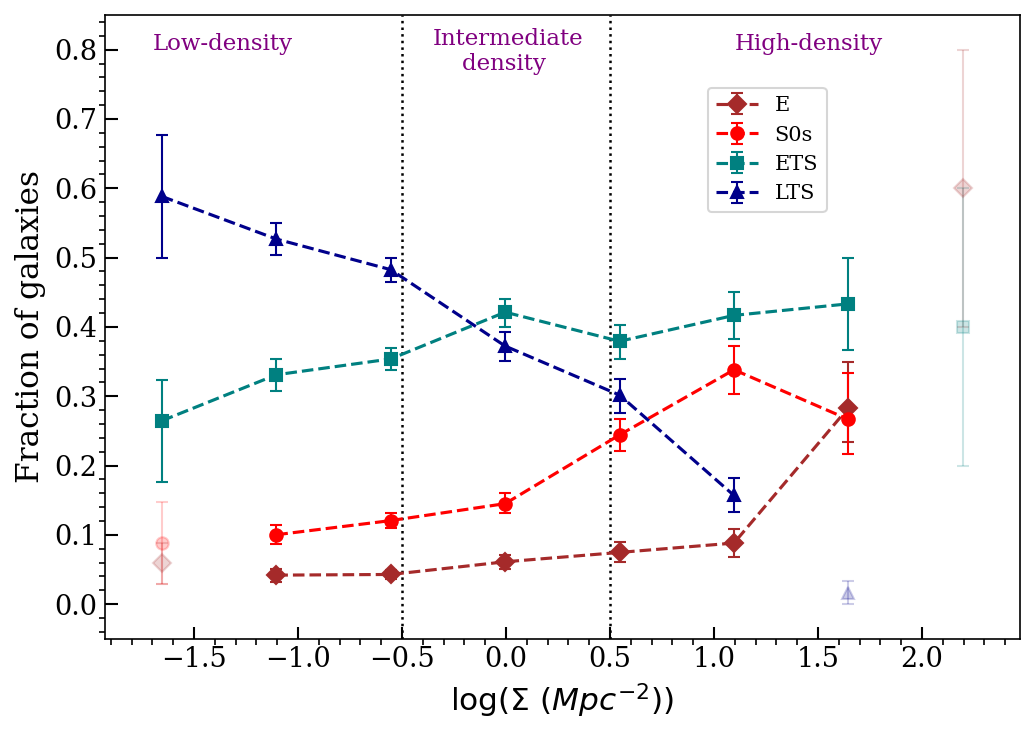}
        \caption{\scriptsize}
        \label{fig:density_all_fraction_low_mass}
    \end{subfigure}
    \begin{subfigure}[t]{0.47\textwidth}
        \includegraphics[width=\textwidth]{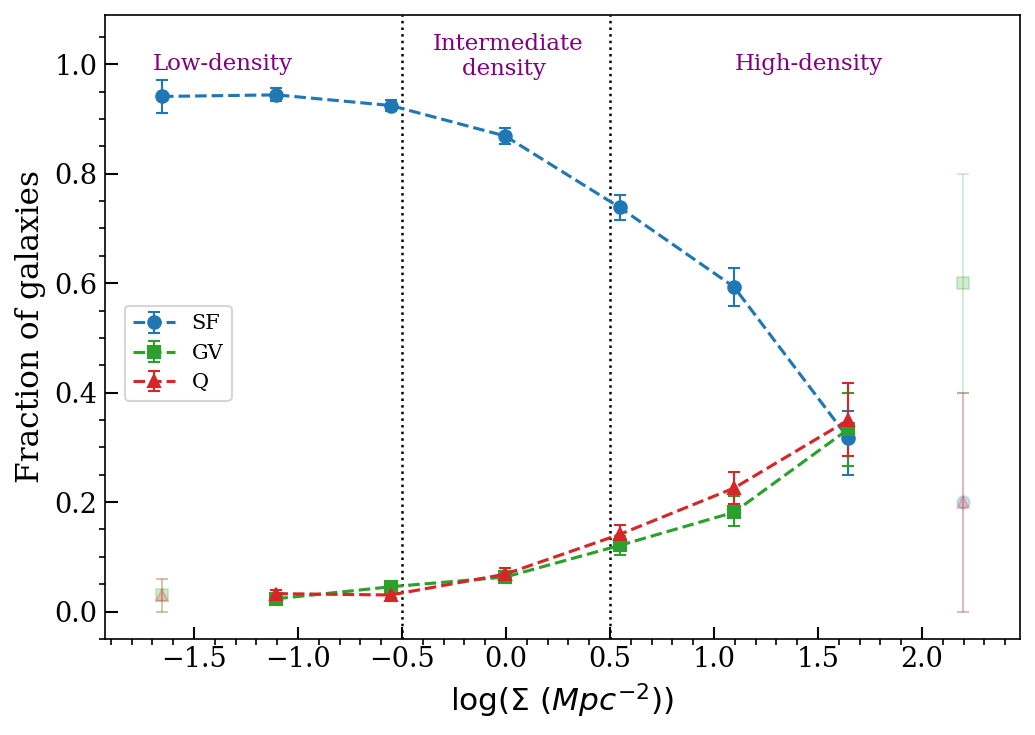}
        \caption{\scriptsize}
        \label{fig:den_low_mass}
    \end{subfigure}
    \caption{Distributions of the 2,515 low-mass galaxies (log(M${\star}$/M${\odot}$) $<$ 10), separated by morphology: ellipticals (E; brown), lenticulars (S0s; red), early-type spirals (ETS; teal), and late-type spirals (LTS; blue).  (a) Morphological composition of the low-mass sample, with E (156), S0s (403), ETS (942), and LTS (1014).  (b) Distribution of local environmental density, with 977 galaxies in low-density, 1,070 in intermediate-density, and 468 in high-density regions.  (c) Distributions of specific star formation rate (sSFR) for each morphological class. (d) Morphological fractions across the star-forming (SF), green valley (GV), and quenched (Q) regions.  (e) Morphological fractions as a function of local environmental density (bin width = 0.55~Mpc$^{-2}$).  (f) Fractions of SF, GV, and Q galaxies as a function of local environmental density (bin width = 0.55~Mpc$^{-2}$). Faded points indicate bins containing fewer than five galaxies. Error bars are estimated using bootstrap resampling.}
    \label{fig:low_mass}
\end{figure*}

The low-mass galaxy sample comprises 2,515 galaxies and is dominated by LTS, as shown in Figure~\ref{fig:morphological_all_histogram_low_mass}. The distributions of local environmental density and sSFR for the low-mass sample are shown in Figures~\ref{fig:density_histogram_low_mass} and~\ref{fig:sSFR_all_histogram_low_mass}, respectively. These distributions indicate that most low-mass galaxies reside in low- to intermediate-density environments and are predominantly star-forming, consistent with expectations from previous studies.

\subsubsection{Morphology–sSFR relation}
We begin by examining the connection between morphology and star formation activity in the low-mass galaxy sample.

As summarized in Table~\ref{tab:morph_fraction}, the low- and high-mass samples exhibit statistically distinct distributions, highlighting the strong dependence of star formation activity on stellar mass. For the low-mass sample, Figure~\ref{fig:sSFR_all_histogram_low_mass} shows that LTS and ETS predominantly occupy the  region, whereas E and S0s are more commonly found in the green valley and quenched regions. These trends are consistent with the morphology-dependent star formation behavior reported in earlier studies.

Figure~\ref{fig:morphology_all_fraction_low_mass} further illustrates the fraction of galaxies in each sSFR category as a function of morphology. Star-forming galaxies are dominated by LTS, with the SF fraction increasing toward later spiral subtypes. Such behavior is well established for low-mass galaxies, which are typically disk-dominated, gas-rich, and characterized by young stellar populations \citep{2003MNRAS.341...54K}. These galaxies generally lie along the star-forming main sequence, where star formation rates scale with stellar mass \citep{2017ApJ...851...22M, 2018A&A...619A..27B}.

In contrast, E and S0s dominate the green valley and quenched regions, indicating suppressed or declining star formation activity. This distribution is consistent with the concept of morphological quenching, whereby dynamically hot, bulge-dominated structures stabilize the gas disk and inhibit star formation \citep{2009ApJ...707..250M, 2010ApJ...721..193P}.

\subsubsection{Effect of local environmental}
We next examine how local environmental density influences the properties of low-mass galaxies. Figure~\ref{fig:density_all_fraction_low_mass} shows the variation of morphological fractions with environmental density. LTS are most prevalent in low-density environments, with their fraction decreasing toward higher densities. In contrast, ETS shows a mild increase with density, while S0s and E exhibit a pronounced rise in high-density environments. These trends indicate a strong correlation between galaxy morphology and local environment in the low-mass regime.

Environmental processes are expected to influence star formation activity through mechanisms such as tidal interactions, mergers, and gas stripping. Observational studies have shown that close interactions and mergers can significantly affect star formation in low-mass galaxies, either enhancing or suppressing star formation depending on interaction stage and gas content \citep{2008MNRAS.385.1903L, 2015ApJ...805....2S, 2024A&A...681A...8S, 2025arXiv250714695C}. The increasing fractions of S0s and E galaxies in denser environments are therefore consistent with environmentally mediated suppression of star formation, although the present analysis remains statistical in nature.

To further quantify these trends, Table~\ref{tab:density_mass_fraction} and Figure~\ref{fig:den_low_mass} summarize the distribution of low-mass galaxies across the star-forming, green valley, and quenched regions as a function of environmental density. In all environments, the low-mass population is dominated by star-forming galaxies, with comparatively smaller fractions occupying the green valley and quenched regions. However, the relative contributions of green valley and quenched galaxies increase systematically in high-density environments. Compared to high-mass galaxies (Figure~\ref{fig:den_high_mass}), this transition occurs more gradually, suggesting that low-mass galaxies can sustain star formation for longer timescales, particularly in low- and intermediate-density environments.

These results are consistent with previous studies showing that environmental quenching is effective for low-mass galaxies at $z < 1$ \citep{Kawinwanichakij_2017}, but proceeds more slowly than in massive galaxies \citep{2018ApJ...853..155D, 2022A&A...666A.141M}. The presence of dwarf galaxies within the low-mass sample may play a significant role in shaping these trends. To assess this possibility, we next isolate and analyze the dwarf galaxy population using the same methodological framework.

\subsection{Effect of Stellar Mass: Dwarfs versus Intermediate-Mass Galaxies} \label{sec:comp}

The analysis of dwarf galaxies is essential, given our earlier finding that a large fraction of star-forming galaxies reside in the low-mass regime. To assess whether the observed trends are representative of the entire low-mass population or are influenced by specific subcomponents, we examine the dwarf and intermediate-mass galaxy subsets separately.

The dwarf and intermediate-mass samples consist of 977 and 1538 galaxies, respectively. This enables a direct comparison between the two mass regimes on equal footing, avoiding biases that may arise when combining the full low-mass sample into a single population.

\begin{figure*}[htbp]
    \centering
    \begin{subfigure}[t]{0.47\textwidth}
        \includegraphics[width=\textwidth]{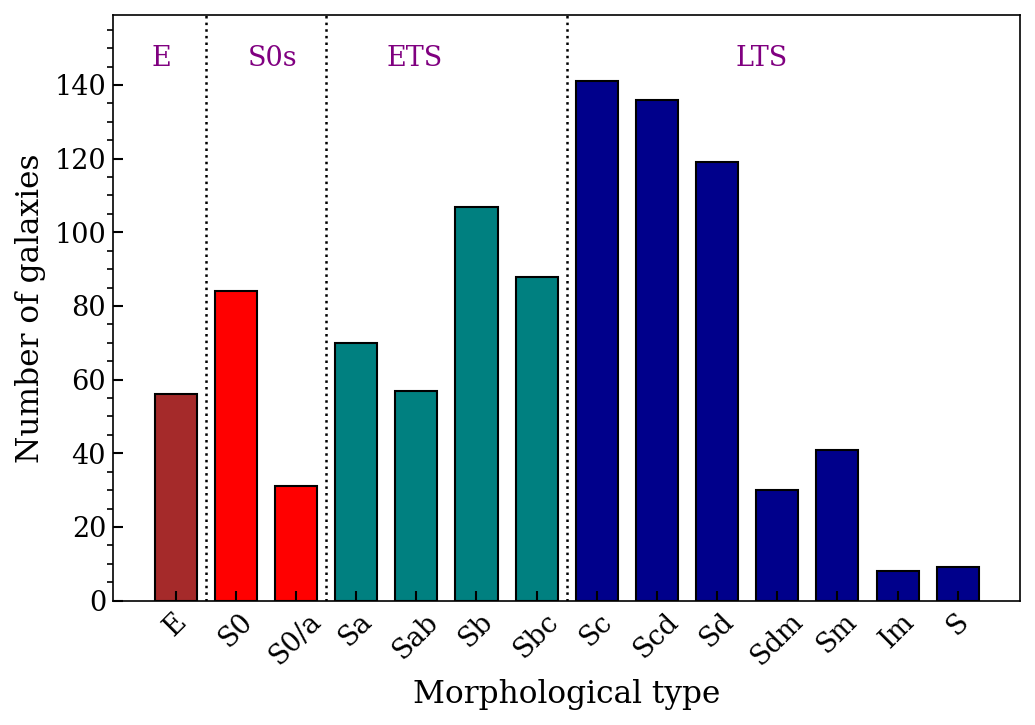}
        \caption{\scriptsize}
        \label{fig:dwarf_a}
    \end{subfigure}
    \begin{subfigure}[t]{0.47\textwidth}
        \includegraphics[width=\textwidth]{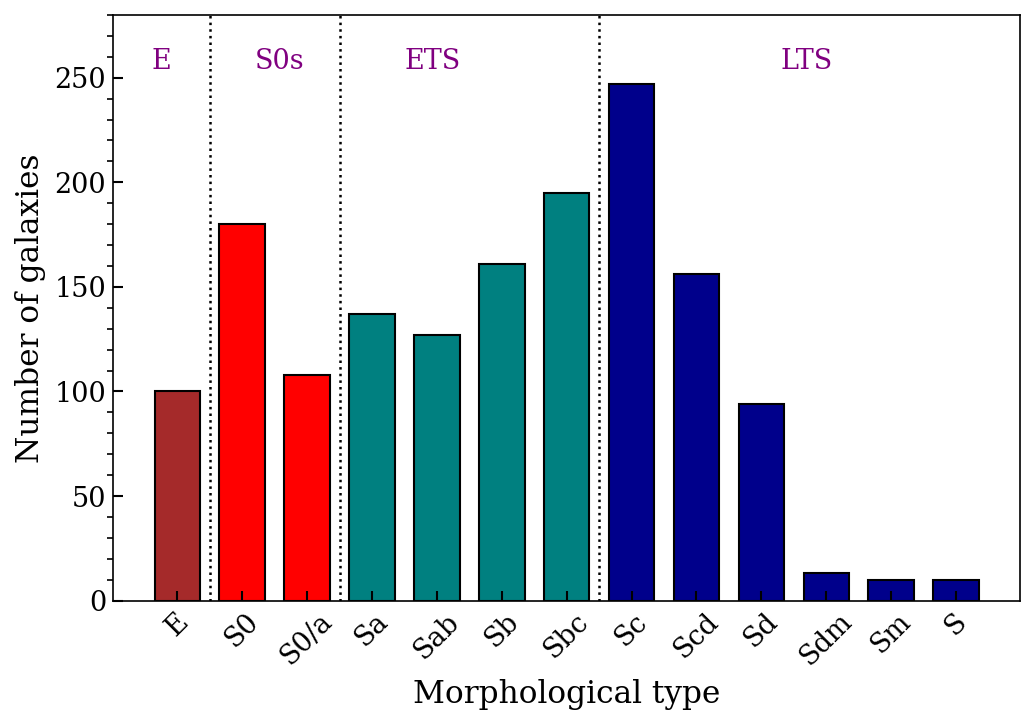}
        \caption{\scriptsize}
        \label{fig:int_mass_a}
    \end{subfigure}
    
    \vspace{0.1cm}
    
    \begin{subfigure}[t]{0.47\textwidth}
        \includegraphics[width=\textwidth]{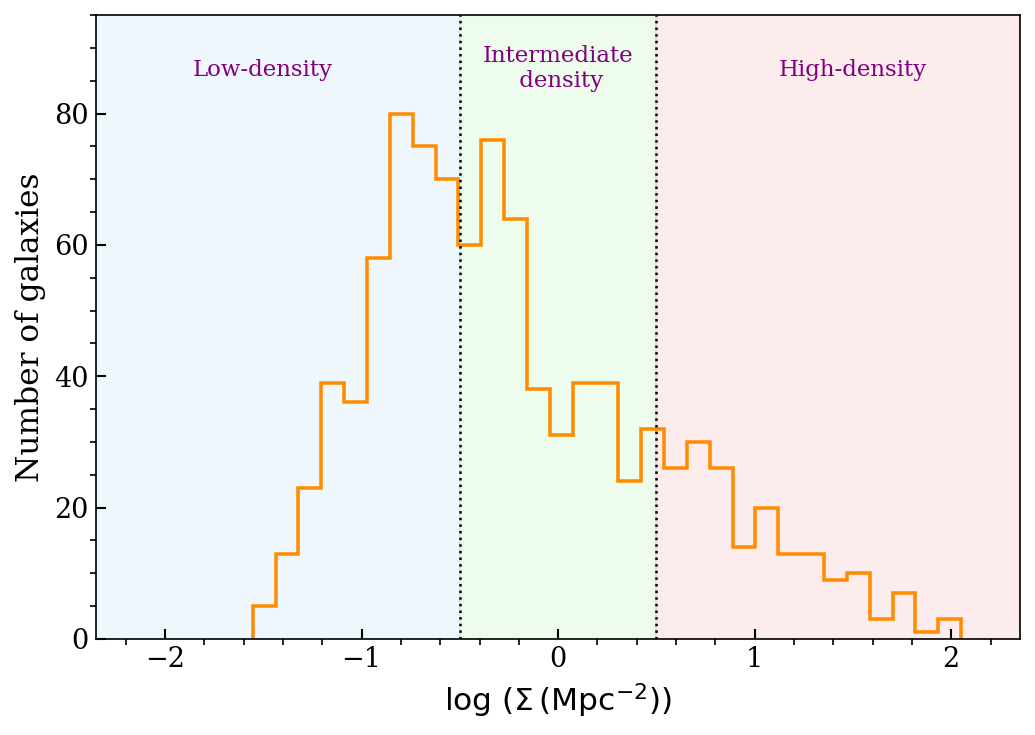}
        \caption{\scriptsize}
        \label{fig:dwarf_b}
    \end{subfigure}
    \begin{subfigure}[t]{0.47\textwidth}
        \includegraphics[width=\textwidth]{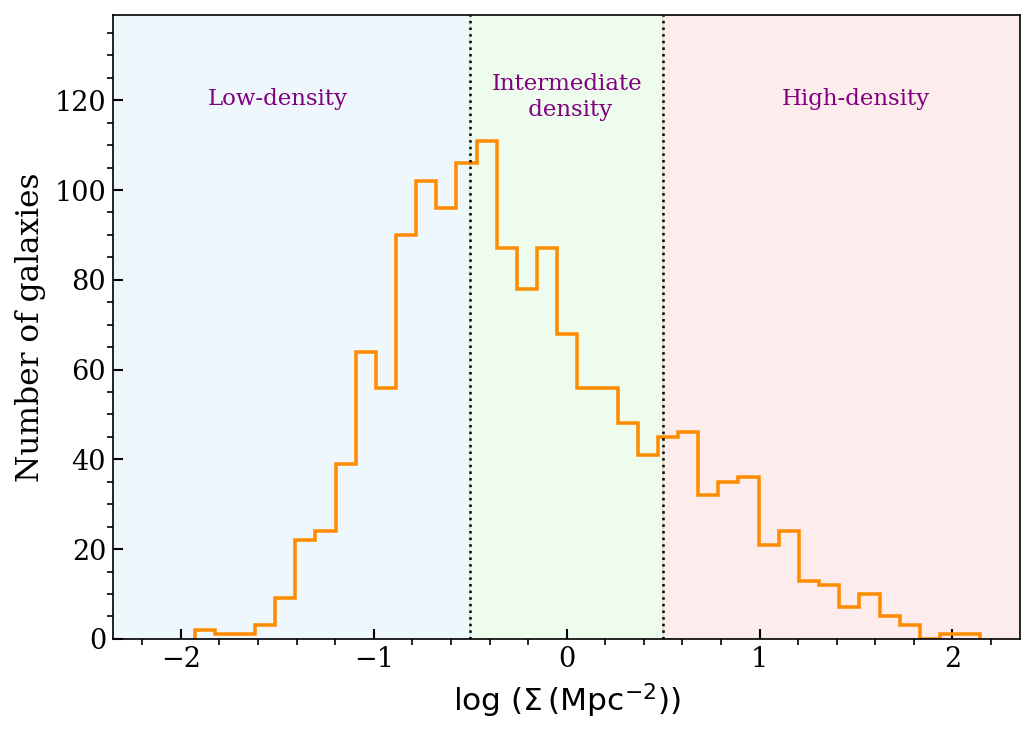}
        \caption{\scriptsize}
        \label{fig:int_mass_b}
    \end{subfigure}
    
    \vspace{0.1cm}
    
    \begin{subfigure}[t]{0.47\textwidth}
        \includegraphics[width=\textwidth]{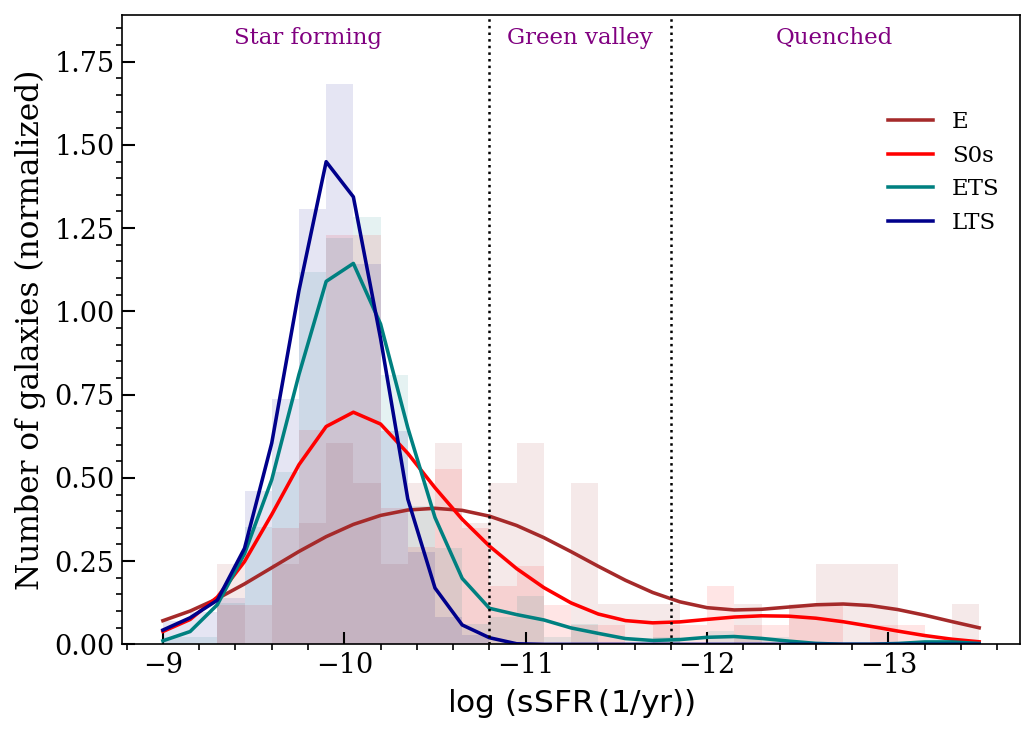}
        \caption{\scriptsize}
        \label{fig:dwarf_c}
    \end{subfigure}
    \begin{subfigure}[t]{0.47\textwidth}
        \includegraphics[width=\textwidth]{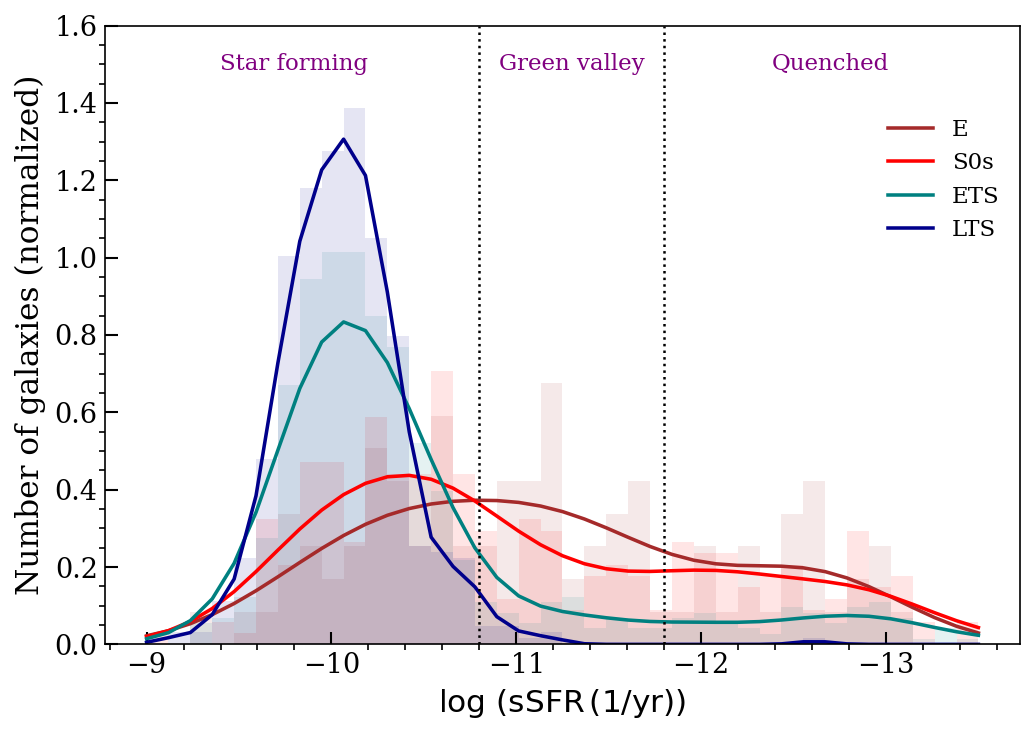}
        \caption{\scriptsize}
        \label{fig:int_mass_c}
    \end{subfigure}
    
    \caption{Distributions of galaxies in two stellar mass ranges. The left column shows galaxies with $\log(M_\star/M_\odot) \leq 9.5$ (977 dwarf galaxies), while the right column shows galaxies with $9.5 < \log(M_\star/M_\odot) < 10.0$ (1,538 intermediate-mass galaxies), separated by morphology: ellipticals (E; brown), lenticulars (S0s; red), early-type spirals (ETS; teal), and late-type spirals (LTS; blue). (a) Morphological composition of the sample. \textit{left:} E (56), S0s (115), ETS (322), and LTS (484). \textit{right:} E (100), S0s (288), ETS (620), and LTS (530). (b) Distribution of local environmental density. For left: 399 galaxies in low-density, 389 in intermediate-density, and 189 in high-density regions. For right: 578 galaxies in low-density, 681 in intermediate-density, and 279 in high-density environments. (c) Distributions of specific star formation rate (sSFR) for each morphological class.}
    \label{fig:mass_comparison}
\end{figure*}

\begin{figure*}[htbp]\ContinuedFloat
    \centering
    \begin{subfigure}[t]{0.47\textwidth}
        \includegraphics[width=\textwidth]{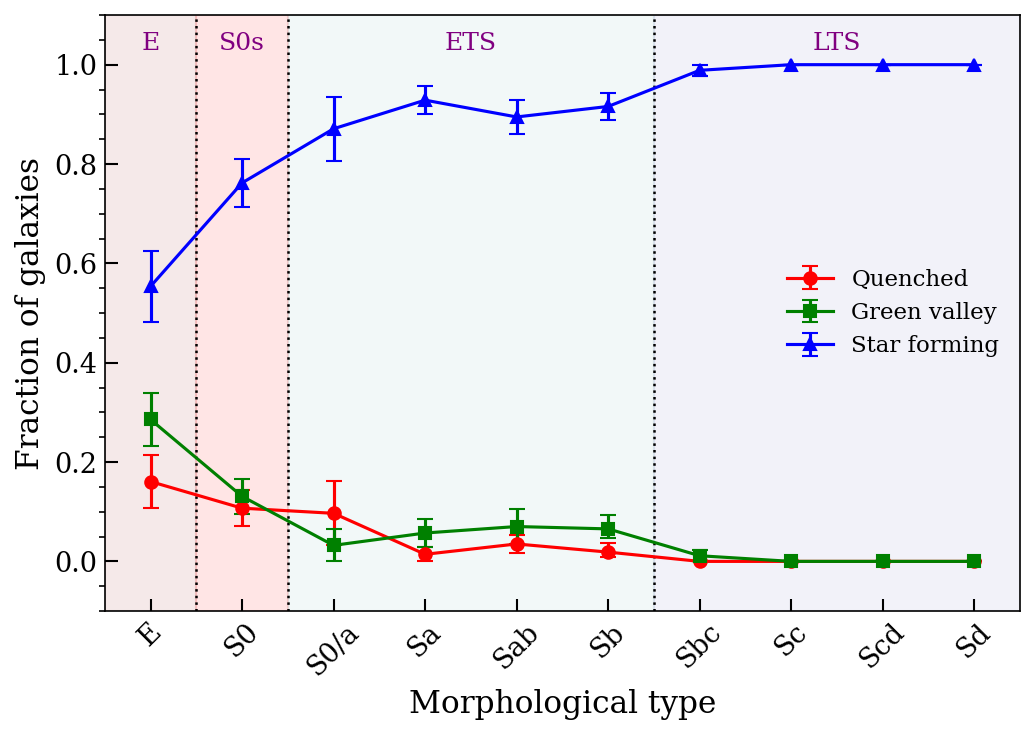}
        \caption{\scriptsize}
        \label{fig:dwarf_d}
    \end{subfigure}
    \begin{subfigure}[t]{0.47\textwidth}
        \includegraphics[width=\textwidth]{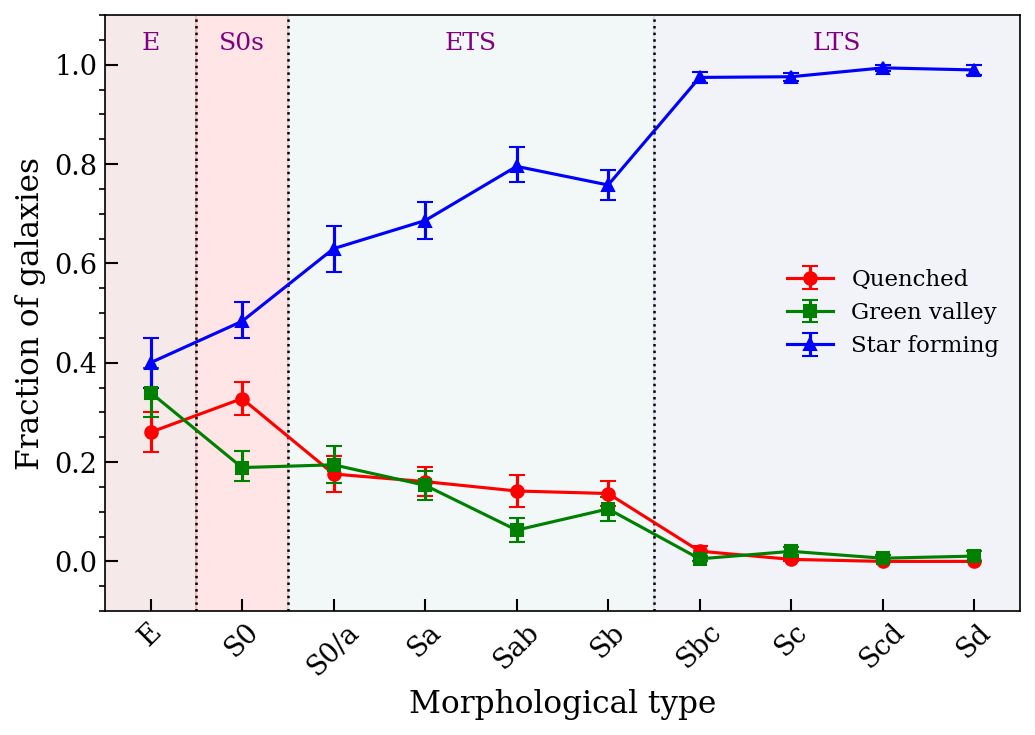}
        \caption{\scriptsize}
        \label{fig:int_mass_d}
    \end{subfigure}
    
    \vspace{0.1cm}
    
    % Row 5
    \begin{subfigure}[t]{0.47\textwidth}
        \includegraphics[width=\textwidth]{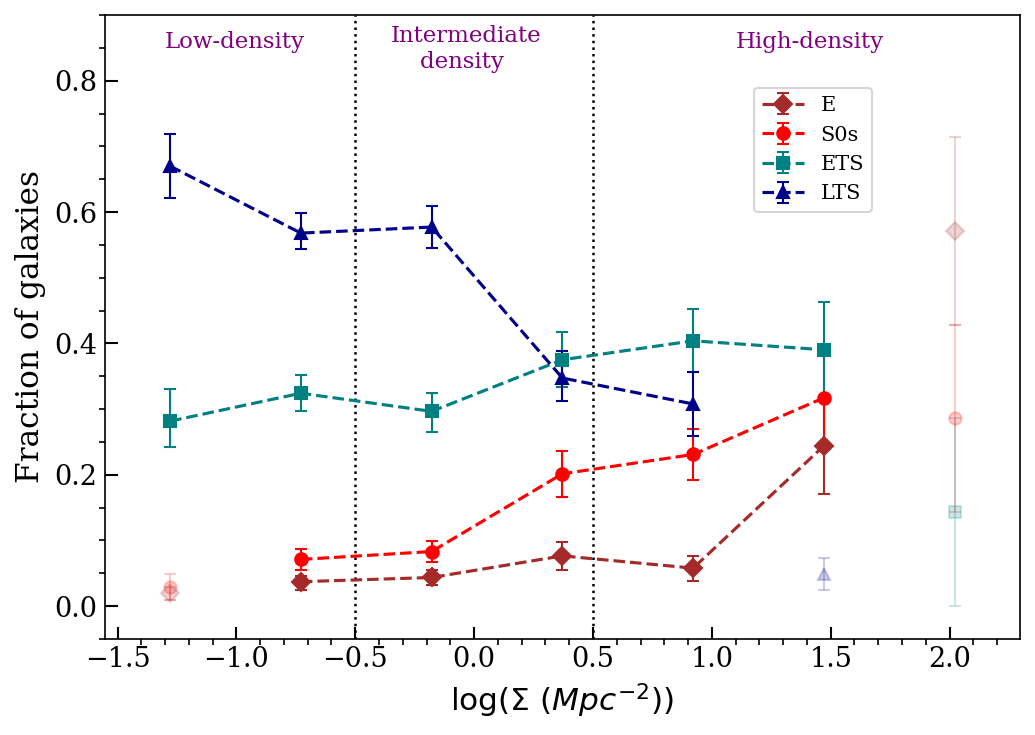}
        \caption{\scriptsize}
        \label{fig:dwarf_e}
    \end{subfigure}
    \begin{subfigure}[t]{0.47\textwidth}
        \includegraphics[width=\textwidth]{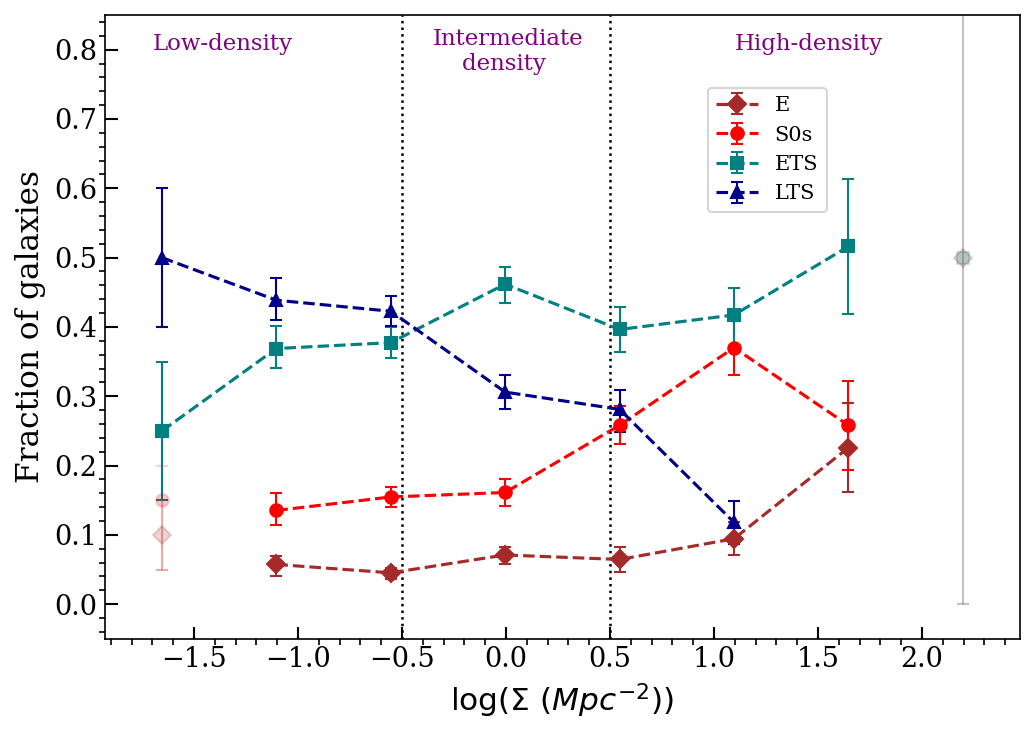}
        \caption{\scriptsize}
        \label{fig:int_mass_e}
    \end{subfigure}
    
    \vspace{0.1cm}
    
    % Row 6
    \begin{subfigure}[t]{0.47\textwidth}
        \includegraphics[width=\textwidth]{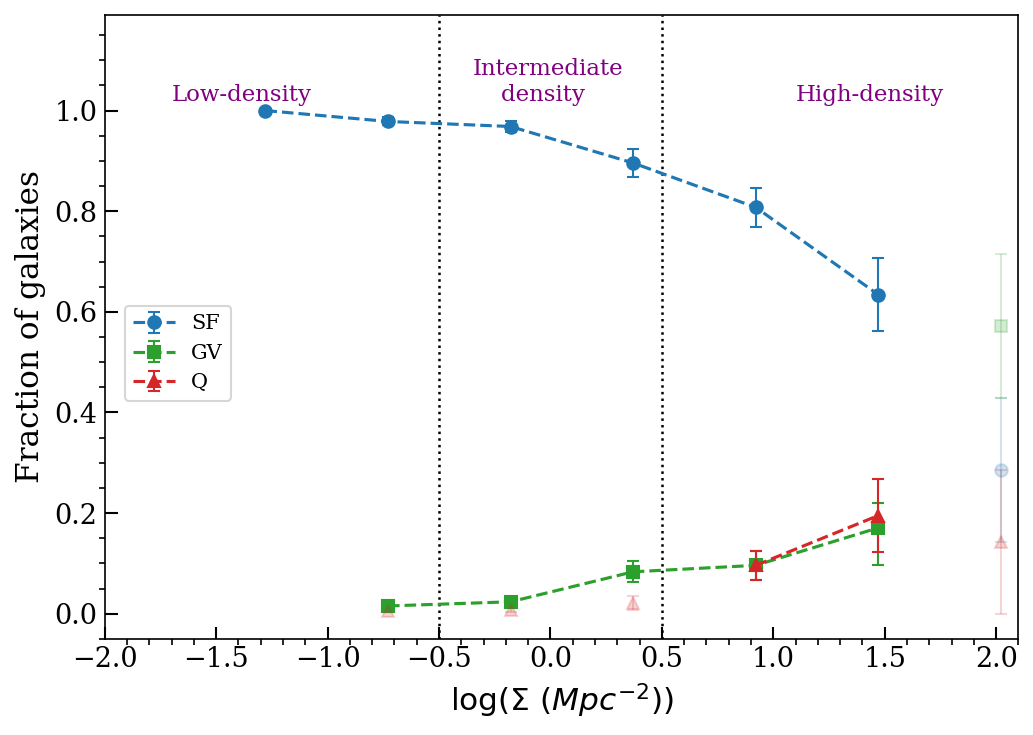}
        \caption{\scriptsize}
        \label{fig:dwarf_f}
    \end{subfigure}
    \begin{subfigure}[t]{0.47\textwidth}
        \includegraphics[width=\textwidth]{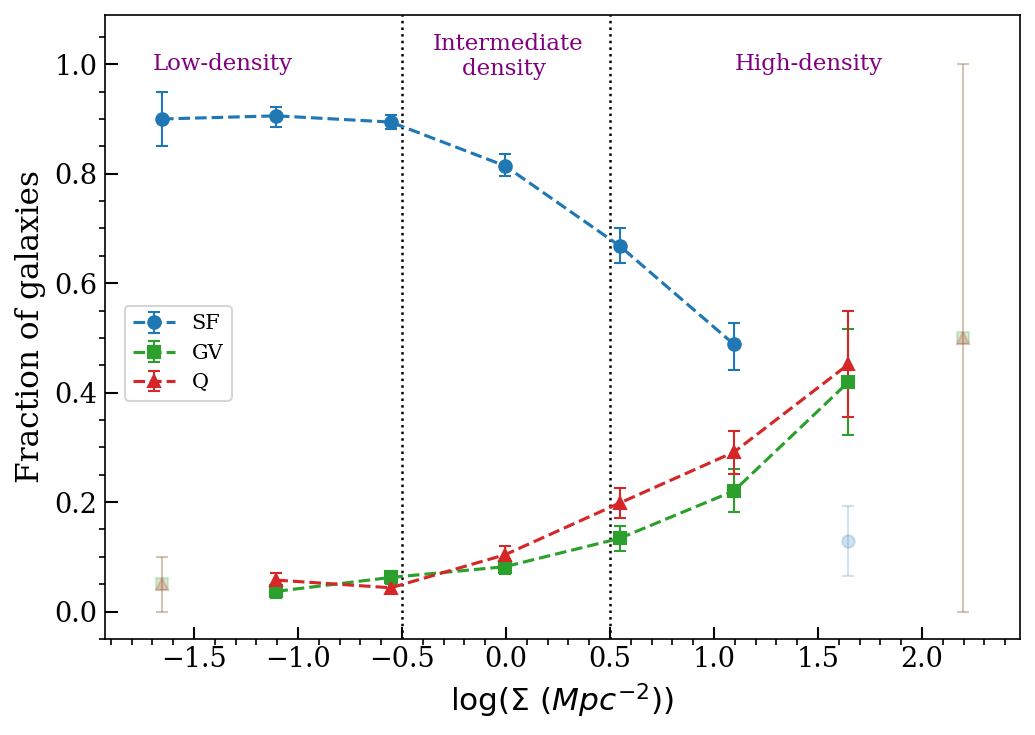}
        \caption{\scriptsize}
        \label{fig:int_mass_f}
    \end{subfigure}
    \caption{Figure~\ref{fig:mass_comparison} (continued). (d) Morphological fractions across the star-forming (SF), green valley (GV), and quenched (Q) regions. (e) Morphological fractions as a function of local environmental density (bin width = 0.55~Mpc$^{-2}$). (f) Fractions of SF, GV, and Q galaxies as a function of local environmental density (bin width = 0.55~Mpc$^{-2}$). Faded points indicate bins containing fewer than five galaxies. Error bars are estimated using bootstrap resampling.}
\end{figure*}

In terms of morphology, Figures~\ref{fig:dwarf_a} and \ref{fig:int_mass_a} show that dwarf galaxies are predominantly composed of LTS, whereas intermediate-mass galaxies are dominated by ETS. Since LTS galaxies are generally more disk-dominated and less centrally concentrated than ETS galaxies, this indicates that dwarf galaxies have a higher prevalence of disk-dominated morphologies, while intermediate-mass galaxies contain a larger fraction of systems with more prominent bulge components.

In terms of environment, Figures~\ref{fig:dwarf_b} and \ref{fig:int_mass_b} show that dwarf galaxies exhibit a bimodal distribution across low- and intermediate-density environments, where most of their population resides. In contrast, intermediate-mass galaxies peak in intermediate-density regions. This suggests that dwarf galaxies are more widely distributed across lower-density environments, whereas intermediate-mass galaxies tend to concentrate in moderately dense environments.

Regarding star-formation activity, Figures~\ref{fig:dwarf_c} and \ref{fig:int_mass_c} show that LTS and ETS galaxies in both dwarf and intermediate-mass regimes are predominantly star-forming. However, E and S0s dwarf galaxies show a larger fraction of star-forming and green valley populations compared to their intermediate-mass counterparts, which display a more substantial population in the quenched region. Figures~\ref{fig:dwarf_d} and \ref{fig:int_mass_d} further support this trend, showing that the morphological fractions of E and S0s in the dwarf regime contribute more to the star-forming population, while intermediate-mass E and S0s are more strongly associated with quenching.

These results indicate that the connection between morphology and star-formation activity differs significantly between dwarf and intermediate-mass galaxies. With increasing environmental density, the fraction of LTS galaxies decreases in both mass regimes, although the decline is more pronounced for dwarf galaxies. In contrast, the fractions of S0s and E galaxies remain relatively constant up to intermediate densities and then increase sharply in high-density environments. ETS dwarf galaxies show a nearly constant behavior across low- and intermediate-density environments, with only a slight increase in high-density regions. However, ETS intermediate-mass galaxies show a gradual increase in fraction from low- to high-density environments (Figures~\ref{fig:dwarf_e} and \ref{fig:int_mass_e}). These trends suggest that environmental processes influence morphological transformation more strongly in intermediate-mass galaxies than in dwarfs.

These patterns become even clearer in Figures~\ref{fig:dwarf_f} and \ref{fig:int_mass_f}, where the fraction of star-forming galaxies decreases toward higher-density environments, while the fractions of green valley and quenched galaxies increase. This trend is particularly pronounced for intermediate-mass galaxies, which show a significantly larger transition and quenched population in dense environments.

In summary, these results indicate that dwarf galaxies are predominantly disk-dominated and actively star-forming across a wide range of environments, whereas intermediate-mass galaxies exhibit stronger environmental dependence, with increasing fractions of quenched and ETS in denser regions. This highlights systematically different evolutionary trends of dwarf and intermediate-mass galaxies, where environmental effects play a more significant role in driving quenching and morphological transformation in the intermediate-mass regime. This suggests that dwarf galaxies likely contribute significantly to the high fraction of star-forming galaxies in the overall low-mass sample. These results indicate a gradual transition in galaxy properties across the low-mass regime, rather than a single homogeneous population.

\subsection{High-mass galaxies}\label{sec:Data Analysis of high-mass galaxies}
While the primary focus of this work is on low-mass galaxies, we briefly examine the high-mass population for comparison. The high-mass sample consists of 4,893 galaxies with $\log (M_{\star}/M_{\odot}) \geq 10$, allowing us to assess whether the trends identified at low mass persist at higher stellar masses.

In the high-mass regime, we recover well-established correlations between morphology, star formation activity, and local environmental density. Early-type galaxies (E and S0s) dominate the quenched population and are preferentially found in denser environments, whereas spiral galaxies primarily occupy the star-forming and green valley regions. Compared to intermediate-mass and dwarf galaxies, the transition between star-forming and quenched states appears to depend more strongly on stellar mass than on environment.

Overall, the behavior of high-mass galaxies is consistent with previous SDSS-based studies \citep[e.g.,][]{2017MNRAS.471.2687B}. A detailed analysis of the morphology–sSFR–environment relations in the high-mass regime is presented in Appendix \ref{appendix:high_mass}. This comparison highlights that the trends commonly attributed to low-mass galaxies are not simple extensions of high-mass behavior, reinforcing the distinct evolutionary role of dwarf galaxies.

\subsection{Comparative analysis of stellar populations across mass regimes}

We investigate the dependence of $\mathrm{D4000}_{R_e}$ on local environmental density and morphology across different stellar mass regimes. This analysis enables us to determine whether environmental effects leave a cumulative imprint on stellar populations and to assess how these imprints differ between dwarf, intermediate-mass, and high-mass galaxies.

\begin{figure}[htbp]
    \centering
    \hspace*{-1cm}
    \includegraphics[width=1.1\columnwidth]{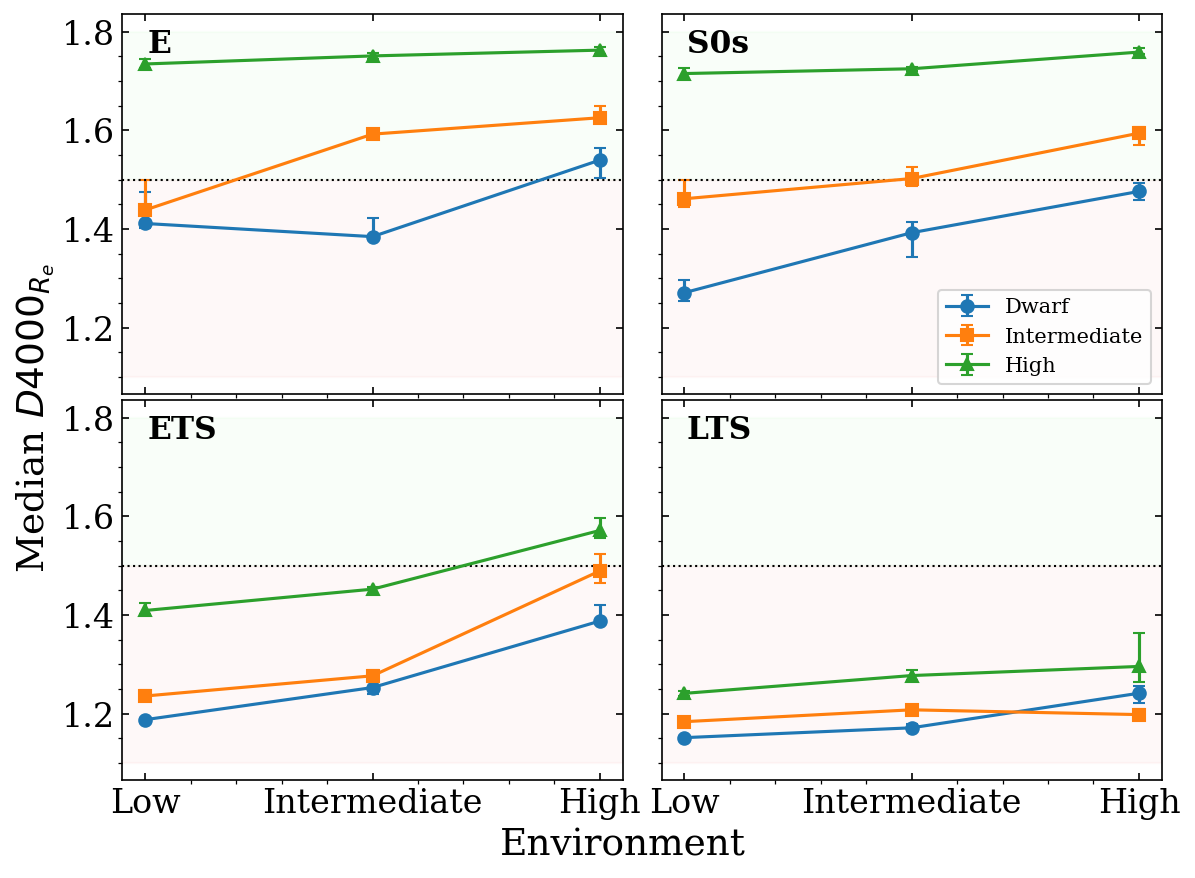}
    \caption{Median D4000$_{R_e}$ index as a function of local environmental density for different morphological types in high-mass, intermediate, and dwarf galaxies. The x-axis represents the environmental density bins corresponding to low-, intermediate-, and high-density regions. Error bars indicate uncertainties estimated from bootstrap resampling. The horizontal dotted line at D4000$_{R_e}=1.5$ separates younger and older stellar populations.}
    \label{fig:Dn4000_scatter_plot}
\end{figure}

The median D4000$_{R_e}$ index shows different trends with local environmental density across the three stellar mass regimes, as illustrated in Figure~\ref{fig:Dn4000_scatter_plot}. The environmental density bins correspond to low-, intermediate-, and high-density regions.

Massive galaxies consistently exhibit higher D4000${R_e}$ values than low-mass galaxies in all environments, indicating generally older stellar populations. For high-mass galaxies, D4000${R_e}$ remains nearly constant from low- to intermediate-density environments, with only a slight increase at the highest densities, primarily for ETS galaxies.

In contrast, intermediate-mass and dwarf galaxies show a stronger increase in D4000${R_e}$ with environmental density. Among E, intermediate-mass galaxies show older stellar populations in intermediate-density regions, while dwarf E host comparatively younger populations. For S0s, the intermediate-mass galaxies lie close to the D4000${R_e}=1.5$ boundary, indicating relatively older populations, whereas dwarf S0s remain younger in low- and intermediate-density environments. Spiral galaxies (both ETS and LTS) generally show younger stellar populations across all environments.

Our result shows that the intermediate-mass population shows a clearer environmental dependence than dwarf galaxies, particularly for E and S0s. This suggests that environmental effects influence the stellar populations of intermediate-mass early-type galaxies more strongly than those of their dwarf counterparts.

\section{Conclusions and Summary} \label{sec:conclusion}
We analyzed a sample of 7,408 galaxies from the MaNGA survey to investigate how star formation, local environmental density, and morphology jointly shape galaxy evolution across a wide stellar mass range. Because galaxies with $\log (M_{\star}/M_{\odot}) < 10$ are often treated as a single low-mass population in the literature, we first examine the full low-mass sample to facilitate comparison with previous studies. We then investigate this regime in greater detail by explicitly separating dwarf galaxies ($\log (M_{\star}/M_{\odot}) \leq 9.5$) from more massive low-mass galaxies (intermediate-mass galaxies; $9.5 < \log (M_{\star}/M_{\odot}) < 10.0$). This approach provides a unified but physically motivated view of how stellar mass and environment regulate galaxy properties.

To quantify the underlying stellar populations, we use the D4000 index as a tracer of long-term stellar population aging. Our analysis reveals a clear bimodality in stellar population properties between dwarf and intermediate-mass galaxies, particularly for early-type galaxies (E and S0s). Stellar mass emerges as the primary driver of galaxy evolution, while local environment acts as a secondary modulator, with its effects being most pronounced in intermediate-mass galaxies.

These results demonstrate that dwarf galaxies, especially early-type dwarfs, host distinctly younger stellar populations than their intermediate-mass counterparts despite sharing similar morphological classifications. This divergence is not apparent among spiral galaxies, where stellar population ages are comparatively insensitive to the dwarf versus non-dwarf distinction. Together, these findings indicate that stellar mass plays a particularly important role in regulating star formation histories in early-type galaxies, while morphology dominates the stellar population properties of late-type galaxies. The combined influence of stellar mass and morphology is therefore essential for interpreting galaxy evolutionary trends in the low-mass regime.

Our main findings can be summarized as follows:

\begin{enumerate}
   \item We confirm a strong mass dependence in galaxy star-formation properties. Dwarf galaxies are predominantly star-forming, intermediate-mass galaxies show a transition between star-forming and quenched populations, and high-mass galaxies are largely quenched. Importantly, the high fraction of star-forming galaxies commonly observed in the overall low-mass population is largely dominated by the dwarf galaxies. This demonstrates that the low-mass regime is characterized by qualitatively similar but quantitatively distinct behavior across stellar mass bins.

  \item We find that morphology further modulates these mass-dependent trends. In the low-mass regime, the fraction of ETS increases with local environmental density, whereas an opposite or weaker dependence is observed at high stellar masses. Low-mass galaxies, especially dwarfs, retain ongoing star formation over longer timescales in low- and intermediate-density environments, consistent with slower and less efficient environmental quenching compared to massive galaxies.
 
  \item Using the D4000 index as a tracer of long-term stellar population aging, we identify a clear bimodality between dwarf and intermediate-mass galaxies within the early-type population. Early-type dwarfs (E and S0s) host systematically younger stellar populations than intermediate-mass galaxies with similar morphologies.This distinction is not apparent among spiral galaxies, whose stellar population ages are largely insensitive to the dwarf versus non-dwarf classification. These results suggest that quenching proceeds more gradually in the dwarf regime, allowing residual or recently suppressed star formation to persist to later epochs.
 
\end{enumerate}

Overall, the observed correlations between morphology, specific star formation rate, stellar population age, and environment highlight the coupled but non-degenerate roles of stellar mass and local conditions in shaping galaxy evolution. Our results demonstrate that the commonly used low-mass galaxy regime is not a homogeneous population. Instead, dwarf galaxies and intermediate-mass galaxies exhibit systematically different evolutionary trends, particularly in their stellar population ages and responses to the environment. This indicates that dwarf galaxies represent a distinct regime in terms of their observed properties and should be treated independently from intermediate-mass galaxies when interpreting trends in the low-mass population. Since dwarf galaxies dominate the star-forming fraction within the low-mass regime, combining them with intermediate-mass galaxies can obscure important physical differences in their evolutionary pathways. This framework provides a useful benchmark for upcoming large-scale integral-field spectroscopic surveys such as SDSS-V, 4MOST, and HECTOR, which will significantly expand the census of intermediate-mass and dwarf galaxies across cosmic time.

\section*{ACKNOWLEDGMENTS}
We thank the anonymous referee for thoughtful comments and suggestions that significantly improved the presentation and focus of this work. Gothai~L. thanks the Indian Institute of Astrophysics (IIA) for supporting her work through a Visiting Student Fellowship at the institute. Funding for the Sloan Digital Sky Survey IV has been provided by the Alfred P. Sloan Foundation, the U.S. Department of Energy Office of Science, and the Participating Institutions. SDSS-IV acknowledges support and resources from the Center for High-Performance Computing at the University of Utah. The SDSS web site is www.sdss.org. SDSS-IV is managed by the Astrophysical Research Consortium for the Participating Institutions of the SDSS Collaboration including the Brazilian Participation Group, the Carnegie Institution for Science, Carnegie Mellon University, the Chilean Participation Group, the French Participation Group, Harvard-Smithsonian Center for Astrophysics, Instituto de Astrof\'isica de Canarias, The Johns Hopkins University, Kavli Institute for the Physics and Mathematics of the Universe (IPMU) / University of Tokyo, Lawrence Berkeley National Laboratory, Leibniz Institut f\"ur Astrophysik Potsdam (AIP), Max-Planck-Institut f\"ur Astronomie (MPIA Heidelberg), Max-Planck-Institut f\"ur Astrophysik (MPA Garching), Max-Planck-Institut f\"ur Extraterrestrische Physik (MPE), National Astronomical Observatories of China, New Mexico State University, New York University, University of Notre Dame, Observat\'ario Nacional / MCTI, The Ohio State University, Pennsylvania State University, Shanghai Astronomical Observatory, United Kingdom Participation Group, Universidad Nacional Aut\'onoma de M\'exico, University of Arizona, University of Colorado Boulder, University of Oxford, University of Portsmouth, University of Utah, University of Virginia, University of Washington, University of Wisconsin, Vanderbilt University, and Yale University. \\

\textit{Softwares}: Topcat \citep{2005ASPC..347...29T}, Matplotlib \citep{Hunter:2007}, NumPy  \citep{harris2020array}, Seaborn \citep{Waskom2021} 

\appendix
\renewcommand{\thefigure}{\Alph{section}\arabic{figure}}
\renewcommand{\thetable}{\Alph{section}\arabic{table}}

\section{Morphology-wise Distribution}
\setcounter{table}{0}

\begin{table*}[htbp]
\centering
\caption{Number of galaxies of different morphological types in the star-forming (SF), green valley (GV), and quenched (Q) regions for low-mass ($\log (M_{\star}/M_{\odot}) < 10$) and high-mass ($\log (M_{\star}/M_{\odot}) \geq 10$) samples.
}
\resizebox{\textwidth}{!}{
\begin{tabularx}{\textwidth}{l *{6}{>{\centering\arraybackslash}X}}
\toprule
\textbf{Morph. Type} & \multicolumn{3}{c}{\textbf{Low Mass(2515)}} & \multicolumn{3}{c}{\textbf{High Mass(4893)}} \\
\cmidrule(lr){2-4} \cmidrule(lr){5-7}
 & \textbf{SF(2130)} & \textbf{GV(188)} & \textbf{Q(197)} & \textbf{SF(1949)} & \textbf{GV(1392)} & \textbf{Q(1552)} \\
\midrule
\textbf{E}    & 71  & 50 & 35 & 35  & 194 & 590 \\
\textbf{S0s}   & 246 & 67 & 90 & 102  & 350 & 600 \\
\textbf{ETS}   & 808 & 63 & 71 & 1320 & 781 & 359 \\
\textbf{LTS}   & 1005 & 8  & 1  & 492 & 67 & 3 \\
\bottomrule
\end{tabularx}
}
\label{tab:morph_fraction}
\end{table*}

\begin{table}[htbp]
\centering
\caption{Number of galaxies of different morphological types—ellipticals (E), lenticulars (S0s), early-type spirals (ETS), and late-type spirals (LTS)—in the star-forming (SF), green valley (GV), and quenched (Q) regions. Results are shown separately for low-mass ($\log (M_{\star}/M_{\odot}) < 10$) and high-mass ($\log (M_{\star}/M_{\odot}) \geq 10$) galaxies across low-, intermediate-, and high-density environments.}
\resizebox{0.85\textwidth}{!}{%
\begin{tabularx}{\textwidth}{l *{9}{>{\centering\arraybackslash}X}}
\toprule
\multicolumn{10}{c}{\textbf{Low Mass}} \\
\midrule
\textbf{Morph. Type} 
& \multicolumn{3}{c}{\textbf{Low (977)}} 
& \multicolumn{3}{c}{\textbf{Intermediate (1070)}} 
& \multicolumn{3}{c}{\textbf{High (468)}} \\
\cmidrule(lr){2-4} \cmidrule(lr){5-7} \cmidrule(lr){8-10}
 & \textbf{SF(913)} & \textbf{GV(36)} & \textbf{Q(28)} 
 & \textbf{SF(935)} & \textbf{GV(69)} & \textbf{Q(66)} 
 & \textbf{SF(282)} & \textbf{GV(83)} & \textbf{Q(103)} \\
\midrule
\textbf{E}   & 26 & 6 & 6 & 32 & 20 & 17 & 13 & 24 & 12 \\
\textbf{S0s} & 80 & 19 & 17 & 95 & 22 & 25 & 71 & 26 & 48 \\
\textbf{ETS} & 319 & 10 & 5 & 374 & 23 & 23 & 115 & 30 & 43 \\
\textbf{LTS} & 488 & 1  & -- & 434 & 4  & 1  & 83 & 3  & -- \\
\midrule
\multicolumn{10}{c}{\textbf{High Mass}} \\
\midrule
\textbf{Morph. Type} 
& \multicolumn{3}{c}{\textbf{Low (1530)}} 
& \multicolumn{3}{c}{\textbf{Intermediate (2376)}} 
& \multicolumn{3}{c}{\textbf{High (987)}} \\
\cmidrule(lr){2-4} \cmidrule(lr){5-7} \cmidrule(lr){8-10}
 & \textbf{SF(796)} & \textbf{GV(389)} & \textbf{Q(345)} 
 & \textbf{SF(961)} & \textbf{GV(722)} & \textbf{Q(693)} 
 & \textbf{SF(192)} & \textbf{GV(281)} & \textbf{Q(514)} \\
\midrule
\textbf{E}   & 10 & 40 & 115 & 21 & 95 & 259 & 4 & 59 & 216 \\
\textbf{S0s} & 43 & 103 & 147 & 44 & 162 & 279 & 15 & 85 & 174 \\
\textbf{ETS} & 516 & 225 & 83 & 665 & 426 & 153 & 139 & 130 & 123 \\
\textbf{LTS} & 227 & 21  & --      & 231 & 39  & 2  & 34 & 7  & 1 \\
\bottomrule
\end{tabularx}
}
\label{tab:density_mass_fraction}
\end{table}

\section{Detailed Discussion on High-mass galaxies} \label{appendix:high_mass}
\setcounter{figure}{0}

\begin{figure}[htbp]
    \centering
    \begin{subfigure}[t]{0.47\textwidth}
        \includegraphics[width=\textwidth]{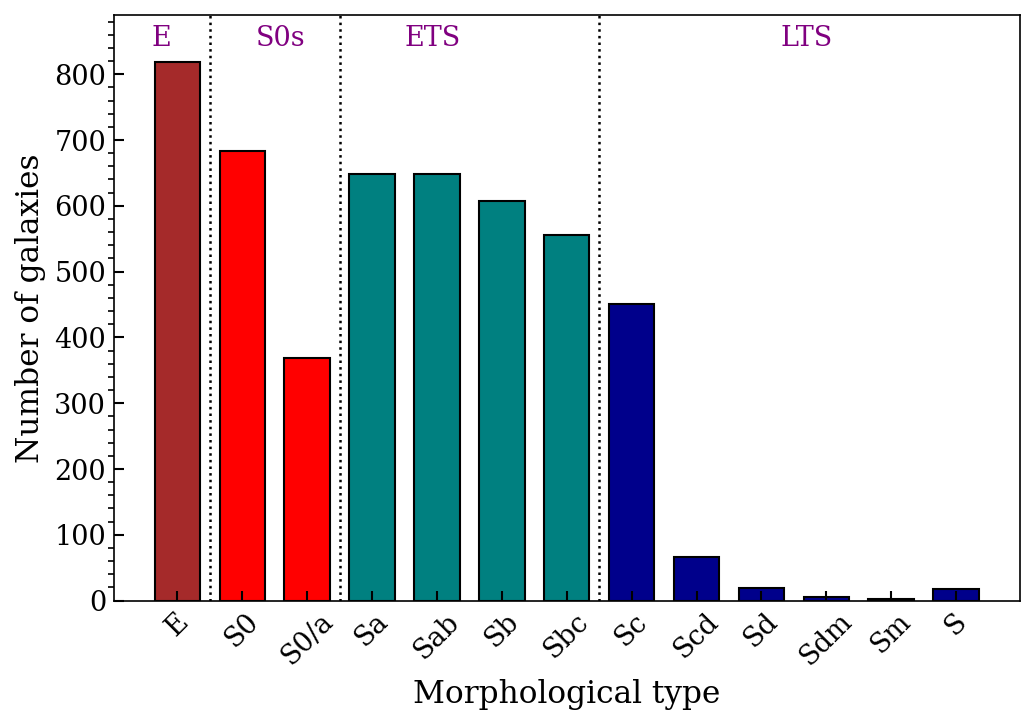} 
        \caption{\scriptsize}
        \label{fig:morphological_all_histogram_high_mass}
    \end{subfigure}
    \begin{subfigure}[t]{0.47\textwidth}
        \includegraphics[width=\textwidth]{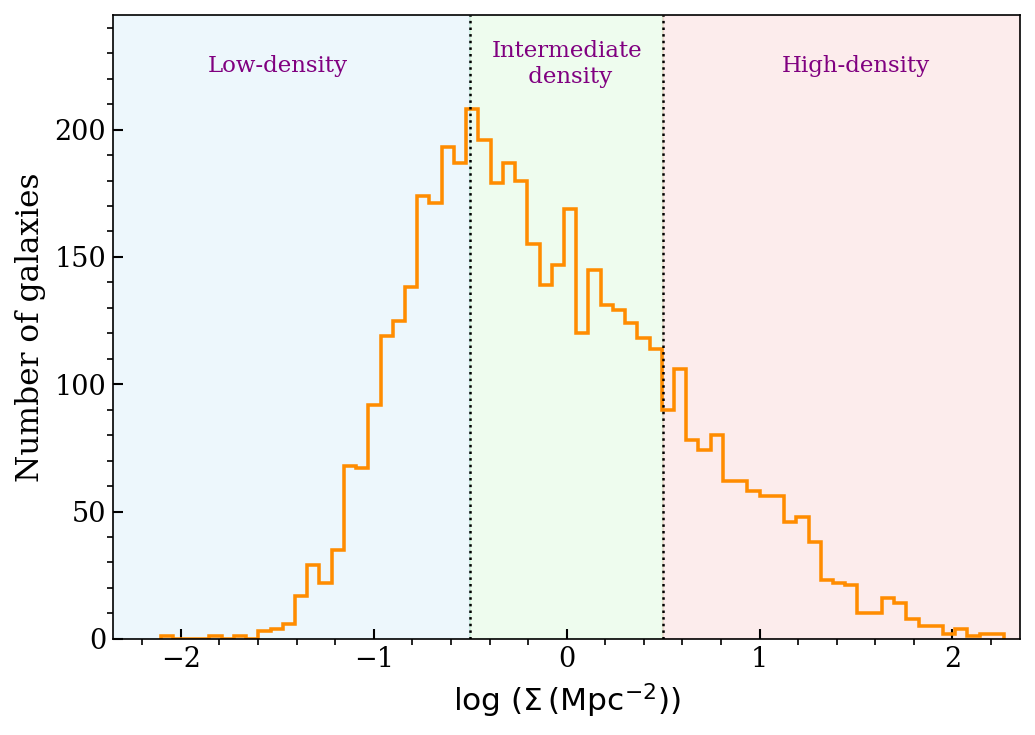} 
        \caption{\scriptsize}
        \label{fig:density_histogram_high_mass}
    \end{subfigure}

    \vspace{0cm} 

    \begin{subfigure}[t]{0.47\textwidth}
        \includegraphics[width=\textwidth]{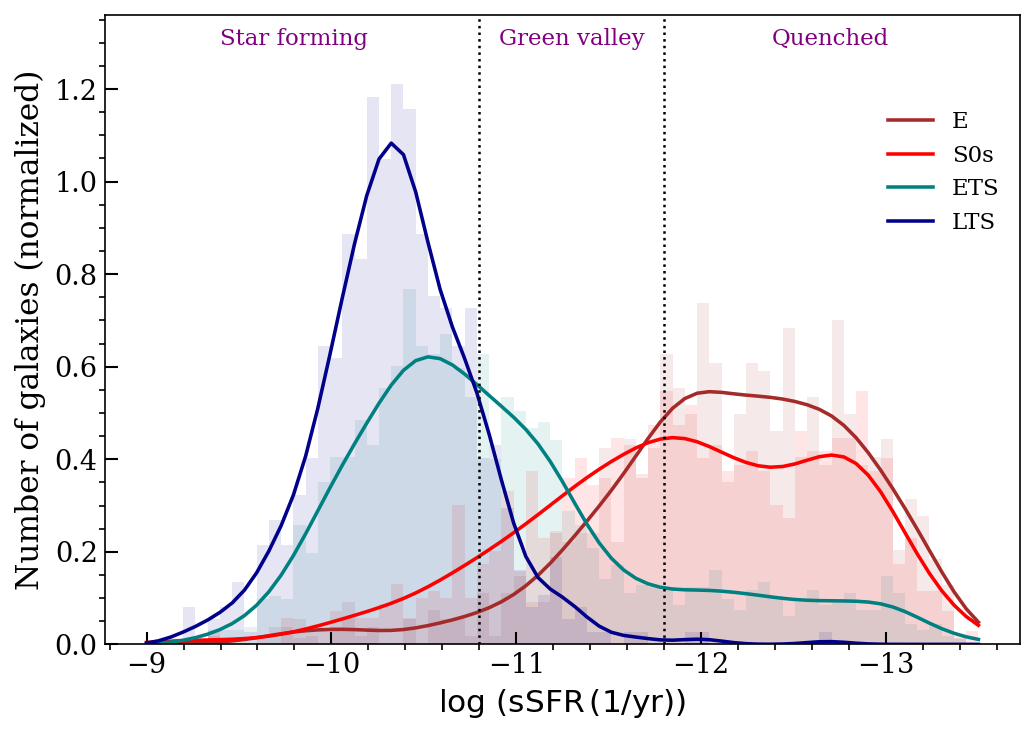}
        \caption{\scriptsize}
        \label{fig:sSFR_all_histogram_high_mass}
    \end{subfigure}
    \begin{subfigure}[t]{0.47\textwidth}
        \includegraphics[width=\textwidth]{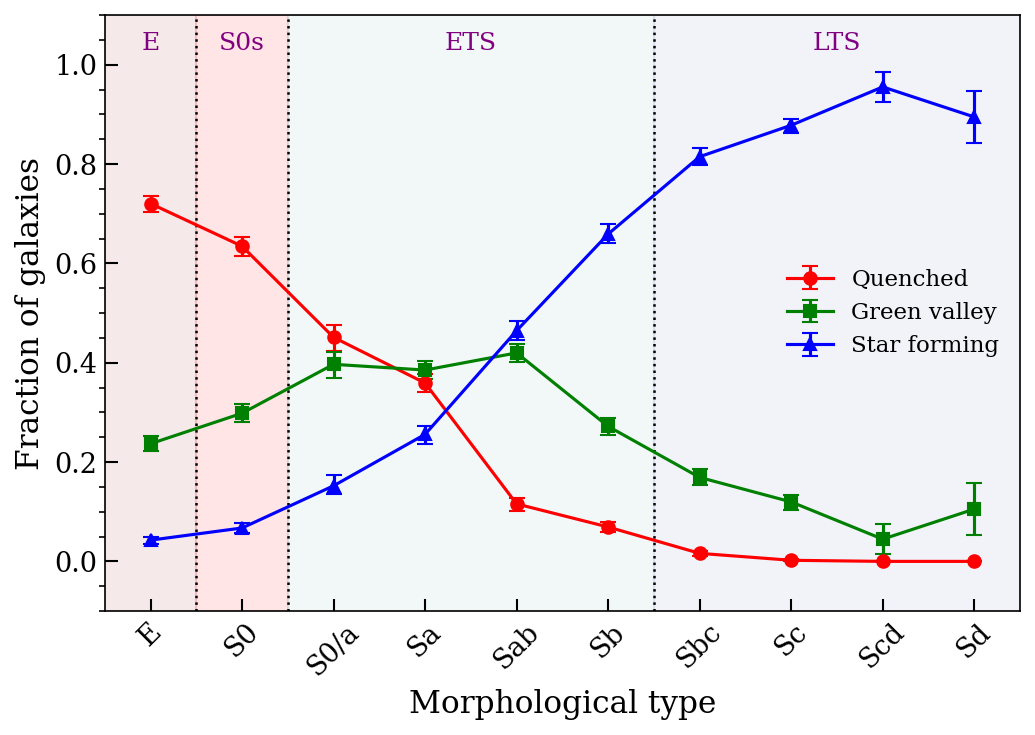}
        \caption{\scriptsize}
        \label{fig:morphology_all_fraction_high_mass}
    \end{subfigure}
    
    \vspace{0cm} % Space between rows
    
    \begin{subfigure}[t]{0.47\textwidth}
        \includegraphics[width=\textwidth]{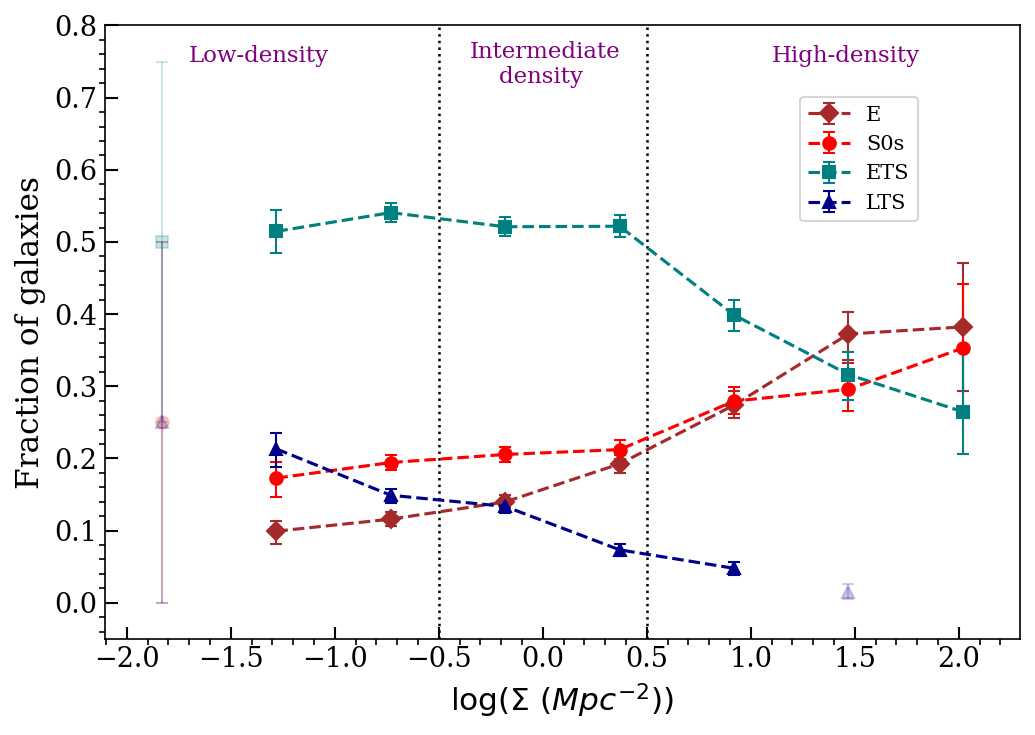}
        \caption{\scriptsize}
        \label{fig:density_all_fraction_high_mass}
    \end{subfigure}
    \begin{subfigure}[t]{0.47\textwidth}
        \includegraphics[width=\textwidth]{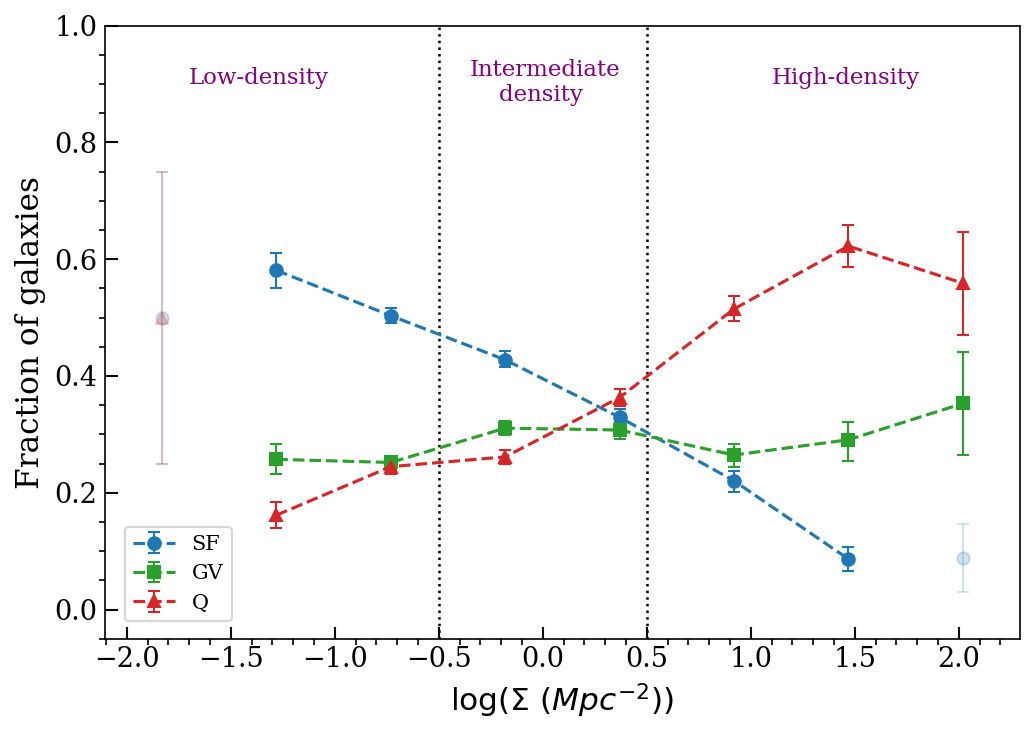}
        \caption{\scriptsize}
        \label{fig:den_high_mass}
    \end{subfigure}
    \caption{Distributions of the 4,893 high-mass galaxies (log(M${\star}$/M${\odot}$) $\geq$ 10), separated by morphology: ellipticals (E; brown), lenticulars (S0s; red), early-type spirals (ETS; teal), and late-type spirals (LTS; blue).  (a) Morphological composition of the high-mass sample, with E (819), S0s (1,052), ETS (2,460), and LTS (562).  (b) Distribution of local environmental density, with 1,530 galaxies in low-density, 2,376 in intermediate-density, and 987 in high-density environments.  (c) Distributions of specific star formation rate (sSFR) for each morphological class. (d) Morphological fractions across the star-forming (SF), green valley (GV), and quenched (Q) regions.  (e) Morphological fractions as a function of local environmental density (bin width = 0.55~Mpc$^{-2}$).  (f) Fractions of SF, GV, and Q galaxies as a function of local environmental density (bin width = 0.55~Mpc$^{-2}$).  Faded points indicate bins containing fewer than five galaxies. Error bars are estimated using bootstrap resampling.}
    \label{5images}
\end{figure}

The high-mass sample comprises 4,893 galaxies. Figures~\ref{fig:morphological_all_histogram_high_mass} and~\ref{fig:density_histogram_high_mass} show the distributions of morphological types and local environmental densities for this sample. Compared to the low-mass galaxies, the high-mass population exhibits a greater fraction of ETS and is more frequently found in intermediate-density environments. These distributions, both in morphology and environment, motivate a detailed examination of how the interplay among sSFR, morphology, and local density evolves in the high-mass domain and are presented here for completeness and comparison with the low-mass, intermediate-mass, and dwarf regimes discussed in the main text.

\subsection{Morphology–sSFR relation}\label{sec:h1}
We examine the relation between galaxy morphology and star formation activity in the high-mass regime for comparison with the low-mass and dwarf populations discussed in the main text. Figure~\ref{fig:sSFR_all_histogram_high_mass} shows the distribution of morphological types across the star-forming, green valley, and quenched regions, with the corresponding fractions summarized in Table~\ref{tab:morph_fraction}.

In the high-mass sample, the star-forming population is dominated by ETS and LTS, with negligible contributions from Es and S0s, indicating that ongoing star formation in massive galaxies is primarily associated with disk-dominated galaxies. ETS galaxies populate both the star-forming and green valley regions, whereas LTS are almost exclusively confined to the star-forming region.

The quenched region is dominated by Es and S0s, with a smaller contribution from ETS and a negligible presence of LTS. The green valley region is primarily populated by ETS and S0s, with moderate contributions from Es. These distributions are consistent with established morphology--star formation trends in massive galaxies.

A closer inspection of Figure~\ref{fig:sSFR_all_histogram_high_mass} reveals a bimodal distribution of S0 galaxies within the quenched region, while ETS and LTS exhibit more continuous distributions. This behavior likely reflects the diversity of evolutionary pathways among massive S0s and is noted here for completeness.

Figure~\ref{fig:morphology_all_fraction_high_mass} presents the distribution of ETS subtypes across the star-forming, green valley, and quenched regions. LTS dominates the star-forming region, while the fraction of ETS increases toward the green valley. Within the quenched population, Sa galaxies constitute the largest fraction among ETS subtypes. These trends are consistent with previous studies and are included here to provide a reference comparison for the low-mass populations.

\subsection{Effect of local environmental}
Building on the results presented in Appendix \ref{sec:h1}, we examine the dependence of galaxy morphology and star formation activity on local environmental density in the high-mass regime. Figures~\ref{fig:density_histogram_high_mass} and~\ref{fig:density_all_fraction_high_mass} show the distributions of morphological types and their fractional contributions across different environmental densities.

The fraction of ETS is highest in low-density environments and remains approximately constant across intermediate densities, followed by a decline in high-density regions. In contrast, LTS exhibit a steady decrease in their fraction with increasing environmental density.

E and S0s display trends opposite to those of spiral galaxies. The fraction of E increases monotonically with environmental density. S0s show a more complex behavior: their fraction remains nearly constant from low- to intermediate-density environments, followed by a modest decrease and a subsequent increase in the highest-density regime. This behavior is consistent with the bimodality noted in Section~\ref{sec:h1}, particularly within the quenched population, and is documented here for completeness.

The distribution of star formation activity across environmental densities is summarized in Table~\ref{tab:density_mass_fraction}. Star-forming galaxies are most prevalent in low-density environments, while quenched galaxies dominate at high densities. Green valley galaxies are more uniformly distributed across all environments. Although the correlation between morphology and star formation remains strong, local environmental density introduces secondary trends that are most evident for E, LTS, and S0s. Notably, the environmental dependence of ETS in the high-mass regime differs from that observed for low-mass galaxies, as discussed in the main text.

Overall, the trends observed for high-mass galaxies are consistent with those reported by \citet{2017MNRAS.471.2687B}, who analyzed a sample of 6,194 galaxies with $\log(M_{\star}/M_{\odot}) > 10$. The agreement between their results and those obtained for the 4,893 high-mass galaxies in the present sample reinforces the robustness of morphology--star formation--environment relations at the high-mass end.

\bibliographystyle{aasjournalv7}
\bibliography{sample701}

\end{document}